%% file: Zprime_arXiv_v2.tex
\documentclass[a4paper,11pt]{article}
\pdfoutput=1 

\usepackage{jheppub} 
\makeatletter
\DeclareRobustCommand*{\bfseries}{%
  \not@math@alphabet\bfseries\mathbf
  \fontseries\bfdefault\selectfont
  \boldmath
}
\makeatother

\usepackage{calc}
\usepackage{rotating}
\usepackage[english]{babel}
\usepackage[utf8]{inputenc}
\usepackage{graphicx}
\usepackage{subfig}
\usepackage{float}
\usepackage{amsmath}
\usepackage{amssymb}
\usepackage{amsthm}
\usepackage{latexsym}
\usepackage{dcolumn}
\usepackage{hyperref}
\input macros.tex

\restylefloat{figure}

\title{Constraints on $Z'$ solutions to the flavor anomalies with trans-Planckian asymptotic safety}

\author{Abhishek Chikkaballi,}
\author{Wojciech Kotlarski,}
\author{Kamila Kowalska,}
\author{Daniele Rizzo}
\author{\\and Enrico Maria Sessolo}

\affiliation{National Centre for Nuclear Research,\\
Pasteura 7, 02-093 Warsaw, Poland}

\emailAdd{abhishek.chikkaballiramalingegowda@ncbj.gov.pl}
\emailAdd{wojciech.kotlarski@ncbj.gov.pl}
\emailAdd{kamila.kowalska@ncbj.gov.pl}
\emailAdd{daniele.rizzo@ncbj.gov.pl}
\emailAdd{enrico.sessolo@ncbj.gov.pl}

\abstract{Motivated by the flavor anomalies in $b\to s$ transitions, we embed minimal models with a $Z'$ gauge boson, 
vector-like fermions, and a singlet scalar in the framework of trans-Planckian asymptotic safety. The presence of a fixed point in the renormalization group flow of the models' parameters leads to predictions for the $\gamma/Z'$ kinetic mixing, 
the New Physics Yukawa couplings, and the quartic couplings of the scalar potential. 
We derive the constraint on the kinetic mixing from the most recent high-mass dilepton resonance searches at the LHC, showing that this bound is often inescapable in this framework, 
unless the U(1) charges conspire to forbid the radiative generation of kinetic mixing. In the latter case, the parameter space consistent with the flavor anomalies can still be probed in depth by direct LHC searches for heavy vector-like quarks and leptons. We derive the current exclusion bounds and projections for future high-luminosity runs.}

\begin{document}
\maketitle

\setcounter{footnote}{0}

\section{Introduction\label{sec:intro}}

In the last few years several ``flavor'' anomalies have appeared in data 
published by the LHCb and other experimental collaborations. The most intriguing and statistically significant of these anomalous measurements involve semileptonic $b\to s$ transitions with 
muons in the final state. In particular, the measurement of the ratios
$R_K=\textrm{BR}(B\to K \mu^+ \mu^-)/\textrm{BR}(B\to K e^+ e^-)$\cite{LHCb:2021trn}, 
including the full Run~I + Run~II data sets,
and $R_{K^{\ast}}=\textrm{BR}(B\to K^{\ast} \mu^+ \mu^-)/\textrm{BR}(B\to K^{\ast} e^+ e^-)$\cite{LHCb:2017avl,Belle:2019oag} each deviates from the Standard Model~(SM) 
prediction approximately at the $3\,\sigma$ level. While strong deviations in these two ``clean'' observables 
would provide a smoking gun for the violation of lepton-flavor universality, 
the overall empirical frame is buttressed by the concurrent emergence of numerous anomalies in
other measurements of branching ratios and angular observables mediated by the $b\to s\, \mu^+ \mu^-$ transition\cite{Aaij:2015dea,Aaij:2015esa,Khachatryan:2015isa,LHCb:2015svh,LHCb:2016ykl,Belle:2016fev,CMS:2017rzx,ATLAS:2018gqc,LHCb:2020lmf,LHCb:2020gog}. A global picture has surfaced, 
pointing to the potential 
presence of New Physics~(NP), 
whose contribution is currently statistically favored compared to the SM prediction alone 
at the level of more than $5\,\sigma$ -- see, \textit{e.g.}, Refs.\cite{DAmico:2017mtc,Ciuchini:2019usw,Alguero:2019ptt,Alok:2019ufo,Datta:2019zca,Aebischer:2019mlg,Kowalska:2019ley,Ciuchini:2020gvn,Hurth:2020ehu,Alda:2020okk,Carvunis:2021jga,Altmannshofer:2021qrr,Geng:2021nhg,Alguero:2021anc} for recent analyses. 

Global effective field theory~(EFT) 
studies of the $b\to s$ anomalous data have confidently shown that some NP states 
ought to be able to generate 
the four-fermion operators $\mathcal{O}_9^{(\prime)\mu},\mathcal{O}_{10}^{(\prime)\mu}$ of the Weak Effective Theory (WET),
with different combinations of the corresponding Wilson coefficients being equivalently favored as long as $C_{9,\textrm{NP}}^{\mu}$ remains large and negative. A handful of simple models were thus identified early on, which could give rise to one or more of these flavor non-diagonal 
vector operators already at the tree level. Popular choices include
leptoquarks (see, \textit{e.g.},
Ref.\cite{London:2021lfn}
for a recent review) and new abelian gauge bosons (early studies include Refs.\cite{Buras:2012jb,Gauld:2013qja,Altmannshofer:2014cfa,Buras:2013dea,Crivellin:2015mga,AristizabalSierra:2015vqb,Allanach:2015gkd,Chiang:2016qov,Bonilla:2017lsq,DiChiara:2017cjq,King:2017anf}).   

Despite the relatively high statistical significance of the observed anomalies and 
the prompt identification of simple NP models that could be responsible for their appearance, one serious obstacle to establishing a targeted strategy for the direct observation of these models 
at the LHC is their lack of predictivity. Flavor experiments, in fact, usually constrain some 
algebraic combination of the models' couplings and typical 
mass scale but the data are not precise enough to disentangle one prediction from the other. In most cases, some assumptions on the ultraviolet~(UV) completion are needed to increase the experimental 
predictivity of the NP model at hand, assumptions that can take the form of boundary conditions for its free parameters.  

In Ref.\cite{Kowalska:2020gie} some of us considered a solution to the $b\to s$ 
flavor anomalies based on a minimal extension of the SM by a single 
leptoquark and completed that simple model in the deep UV with boundary 
conditions based on the assumption of trans-Planckian asymptotic safety (AS). 
The latter is the simple requirement that the renormalization group (RG) flow of all dimensionless couplings of the matter Lagrangian features fixed points 
(either Gaussian or interactive) above the Planck scale, so to be protected 
against the potential appearance 
of Landau poles. The AS ansatz finds its motivation in the large existing body of work dedicated 
to asymptotically safe quantum gravity, where, following the 
development of functional RG techniques\cite{WETTERICH199390,Morris:1993qb}, it was shown\cite{Reuter:1996cp,Lauscher:2001ya,Reuter:2001ag,Manrique:2011jc}
that the quantum fluctuations of the metric field may induce an interactive fixed point 
in the flow of the gravitational couplings. While early findings were limited to the Einstein-Hilbert truncation of the action, 
in subsequent studies the parameter space was successfully extended to include, on the one side, gravitational effective operators of 
increasing mass dimension\cite{Lauscher:2002sq,Litim:2003vp,Codello:2006in,Machado:2007ea,Codello:2008vh,Benedetti:2009rx,Dietz:2012ic,Falls:2013bv,Falls:2014tra}, and on the other, matter-field operators\cite{Robinson:2005fj,Pietrykowski:2006xy,Toms:2007sk,Tang:2008ah,Toms:2008dq,Rodigast:2009zj,Zanusso:2009bs,Daum:2009dn,Daum:2010bc,Folkerts:2011jz,Oda:2015sma,Eichhorn:2016esv,Christiansen:2017gtg,Hamada:2017rvn,Christiansen:2017cxa,Eichhorn:2017eht}, with the ultimate goal of eventually being able to prove the non-perturbative renormalizability of the full system of gravity and matter. 

The endowment of a matter theory like the leptoquark model in Ref.\cite{Kowalska:2020gie} with boundary conditions derived from the presence of interactive UV fixed points bears important consequences for the predictivity of the theory itself. The actual number of free parameters is in fact restricted 
in theory space to 
the number of relevant directions around the fixed point. 
Irrelevant directions of the trans-Planckian flow near
the fixed point lead, on the other hand, to specific predictions in the infrared~(IR). 
In the SM embedded in trans-Planckian AS, for example, irrelevant parameters include the Higgs quartic coupling\cite{Shaposhnikov:2009pv,Eichhorn:2017als,Kwapisz:2019wrl,Eichhorn:2021tsx}, the hypercharge gauge coupling\cite{Harst:2011zx,Christiansen:2017gtg,Eichhorn:2017lry}, 
and the top Yukawa coupling\cite{Eichhorn:2017ylw}.  
In the leptoquark model mentioned above, the size and relative sign of the NP Yukawa couplings
emerge as predictions from the fixed-point analysis. Their IR 
values may then be plugged into the Wilson coefficients of the EFT to extract a fairly precise expectation for 
the mass range of the leptoquark (at $4-7\tev$\cite{Kowalska:2020gie}), 
as well as a number of complementary signatures in $D$ and $K$ meson decay.

Encouraged by the sharp indications we obtained for the observational properties of the leptoquark 
and by the equally keen predictions that have been extracted in other NP scenarios
with asymptotically safe boundary
conditions\cite{Wang:2015sxe,Grabowski:2018fjj,Reichert:2019car,Domenech:2020yjf,Kowalska:2020zve,Kowalska:2022ypk}, 
we perform in this work a trans-Planckian fixed-point analysis of minimal SM extensions 
with a new abelian gauge boson $Z'$, which can lead 
to a solution for the flavor anomalies.\footnote{The phenomenology of some U(1) 
extensions of the SM giving a solution to the flavor anomalies and being, at the same time, 
constrained in an asymptotically safe framework \textit{not} based on quantum gravity was recently investigated in Ref.\cite{Bause:2021prv}. In Ref.\cite{Boos:2022jvc} trans-Planckian AS inspired from quantum gravity was applied to a SM extension with gauged baryon number, without direct connection to the flavor anomalies.} 
Because the $Z'$ is expected to be massive and to couple to the $b$-$s$ current,
the Lagrangians we consider feature Yukawa interactions of a new scalar field $S$ 
with the SM and vector-like (VL) heavy fermions with color charge. 
The scalar's vacuum expectation value (vev) breaks the abelian gauge symmetry 
giving mass to the $Z'$ and the mixing of VL and SM quarks generates the flavor 
non-diagonal gauge coupling. Moreover, because of the presence of VL fermions charged under both the new abelian and the SM gauge groups, 
the $Z'$ is subject to kinetic mixing with the SM photon and $Z$ boson.

We would like to emphasize that the goal of this work is not that 
of addressing the numerous theoretical issues 
still in need of a solution before a consistent theory of asymptotically 
safe quantum gravity is completed. Much effort in this direction is currently undertaken in the community -- see, \textit{e.g.}, Refs.\cite{Donoghue:2019clr,Bonanno:2020bil}. 
Rather, we want to derive here the observational signatures 
that would emerge if one were able to embed consistently the low-energy Lagrangian 
in the framework of AS gravity. 
In particular, for the scenarios introduced above, 
we will show 
that the trans-Planckian fixed point analysis allows one to obtain precise predictions
for the value of the Yukawa couplings, the abelian kinetic mixing, and the Higgs/$S$ mixing.
Plugging the resulting couplings in the Wilson coefficients 
emerging as favored in EFT analyses, 
one further obtains indications for the NP masses. We will assess the viability of our predictions by performing a 
detailed phenomenological analysis of the existing LHC constraints applying to these models 
and other similar scenarios
and we will derive the prospects for their direct detection
 with future increases in luminosity. Incidentally, we derive in the process 
the most up-to-date mass-dependent bound 
on the kinetic mixing from searches for a heavy $Z'$ at the LHC 13~TeV, updating the 7-TeV result of Ref.\cite{Jaeckel:2012yz}.  

The paper is organized as follows. In \refsec{sec:models} we introduce NP models with an extra U(1) symmetry and VL fermions in the context of the flavor anomalies and recall their general structure. In \refsec{sec:fpan} we briefly review the main concepts of AS and perform a full fixed-point analysis of the scenarios introduced in \refsec{sec:models}. The resulting predictions for the low-scale physics and the related phenomenology are discussed in \refsec{sec:constr}, where we also derive the most recent LHC constraints applying to our models. We summarize our findings and conclude in \refsec{sec:summary}. Appendices feature, respectively, the explicit forms of the rotation matrices to the physical basis, the form of the renormalization group equations (RGEs) in the presence of two U(1) gauge factors, the one-loop beta functions of the gauge, Yukawa, and quartic couplings in the scenarios considered in this study, and a specific example of an extension of our models with dark matter and neutrino masses.

\section{Minimal $Z^\prime$ models for the flavor anomalies}\label{sec:models}

In model-independent global analyses
the anomalous flavor data is confronted with the operators of  
an effective Hamiltonian for the $b \to s l^+l^-$ transition (where $l=e,\mu,\tau$). 
One writes 
\be \label{heff}
\mathcal{H}_{\textrm{eff}}=-\frac{4G_F}{\sqrt{2}}V_{tb}V_{ts}^* \sum_{i,l}(C_i^l O_i^l + C_i^{' l} O_i^{' l}) + \textrm{H.c.}\,,
\ee
where $G_F$ is the Fermi constant and $V_{tb}$, $V_{ts}$ are elements of the Cabibbo-Kobayashi-Maskawa (CKM) matrix. The combination of Wilson coefficients $C_i^{(\prime)l}$ with the four-fermion dimension-six interaction operators $O_i^{(\prime) l}$, invariant under the SU(3)$_{\textrm{c}}\times$U(1)$_{\textrm{em}}$ 
gauge group, effectively describes the effects of short-distance physics within the low-energy physics.

The flavor anomalies strongly hint at NP emerging
in the coefficients of a few semi-leptonic operators with muons in the final state,
\bea
O_9^{(\prime)\mu}&=&\frac{\alpha_{\textrm{em}}}{4\pi}\left(\bar{s}\gamma^\rho P_{L(R)} b\right)\left(\bar{\mu}\gamma_\rho \mu\right),\nonumber\\ 
O_{10}^{(\prime)\mu}&=&\frac{\alpha_{\textrm{em}}}{4\pi}\left(\bar{s}\gamma^\rho P_{L(R)} b\right)\left(\bar{\mu}\gamma_\rho\gamma_5 \mu\right),
\eea
where $\alpha_{\textrm{em}}$ is the fine-structure constant in the Thomson limit and 
$P_{L,R}=(1\mp \gamma_5)/2$ are the standard chiral projectors. We limit 
ourselves in this work to NP models that can give rise to the 
left-chiral (unprimed) pair of Wilson coefficients $C^{\mu}_{9,\textrm{NP}}$, $C^{\mu}_{10,\textrm{NP}}$,   
and to CP-conserving NP effects (real Wilson coefficients). Global analyses have shown that the fit to the flavor anomalies does not improve
significantly by adding the right-chiral components or considering their imaginary part. 

The 2$\sigma$ region consistent with the flavor anomalies  reads\cite{Altmannshofer:2021qrr}
\be\label{eq:bound1}
-1.03\leq C_{9,\textrm{NP}}^{\mu}\leq -0.43
\ee
when $C_{10,\textrm{NP}}^{\mu}=0$, and
\be\label{eq:bound2}
-0.53 \leq C_{9,\textrm{NP}}^{\mu}\left(=-C_{10,\textrm{NP}}^{\mu}\right)\leq -0.25
\ee
when $C_{9,\textrm{NP}}^{\mu}=-C_{10,\textrm{NP}}^{\mu}$. The parameter space allowed in the case when $C_{9,\textrm{NP}}^{\mu}$ and $C_{10,\textrm{NP}}^{\mu}$ are independent objects can be found, \textit{e.g.}, in Ref.\cite{Altmannshofer:2021qrr}.
\bigskip

The Hamiltonian of \refeq{heff} can be completed in the UV
with new massive states above the electroweak symmetry-breaking (EWSB) scale. We focus in this work on scenarios with a 
$Z'$ gauge boson. The most generic Lagrangian parameterizing $Z'$ interactions
with the $b$-$s$ current and the muons 
is given by
\be
\mathcal{L}\supset  Z'_{\rho}\left[\left(g_L^{sb}\, \bar{s}\gamma^{\rho}P_L\, b+ g_R^{sb}\, \bar{s}\gamma^{\rho} P_R\, b  + \textrm{H.c.}\right) +  g_L^{\mu\mu}\,\bar{\mu}\gamma^{\rho} P_L\,\mu+g^{\mu\mu}_{R}\,\bar{\mu}\gamma^{\rho}P_R\,\mu\right],\label{mod_ind_Zp}
\ee
in terms of \textit{a priori} free effective couplings 
$g_{L,R}^{sb}$, $g_{L,R}^{\mu\mu}$.
The Wilson coefficients of the WET 
can be then expressed in terms of the effective couplings as
\be\label{Wils_generic}
C_{9,\textrm{NP}}^{\mu}=-2\,\frac{g_{L}^{sb} g^{\mu\mu}_V}{V_{tb}V_{ts}^{\ast}}\left(\frac{\Lambda_v}{m_{Z'}}\right)^2,   \qquad C_{10,\textrm{NP}}^{\mu}=-2\,\frac{g_{L}^{sb} g_A^{\mu\mu}}{V_{tb}V_{ts}^{\ast}}\left(\frac{\Lambda_v}{m_{Z'}}\right)^2, 
\ee
where $g_V^{\mu\mu}\equiv (g^{\mu\mu}_{R}+g^{\mu\mu}_{L})/2$, $g_A^{\mu\mu}\equiv (g^{\mu\mu}_{R}-g^{\mu\mu}_{L})/2$, $m_{Z'}$ 
indicates the mass of the $Z'$ boson, $V_{tb}=0.999$, $V_{ts}=-0.0404$, and
\be
\Lambda_v=\left(\frac{\pi}{\sqrt{2}G_F\alpha_{\textrm{em}}}\right)^{1/2}\approx 4.94\tev
\ee
defines the typical effective scale of the NP states. Note that to simplify the notation we will 
drop the subscript ``NP'' from the Wilson coefficients  $C^{\mu}_9$, $C^{\mu}_{10}$ hereafter.

\subsection{Quark sector couplings\label{sec:quark}}

In order to render the model renormalizable and gauge-invariant, we extend the gauge symmetry of the SM by an additional abelian gauge factor, U(1)$_X$, with gauge coupling denoted by $g_X$. 
The couplings of the $Z'$ boson with gauge eigenstates are by definition flavor-conserving, therefore an additional structure is required to generate the flavor-violating couplings $g_{L}^{sb}$ and $g_{R}^{sb}$. One minimal solution invokes extending the particle content by a singlet 
scalar field $S$ and a VL pair of left-chiral Weyl spinor doublets, $Q=(U_L,D_L)^T$ and $Q'=(U_R,D_R)$, 
collectively charged under \smgaux\ as
\be
S:(\mathbf{1},\mathbf{1},0,Q_S)\,,
\ee
\be\label{eq:VLquarks}
Q:(\mathbf{3},\mathbf{2},1/6,Q_S) \qquad Q':(\mathbf{\bar{3}},\mathbf{\bar{2}},-1/6,-Q_S)\,,
\ee
where $Q_S$ is the U(1)$_X$ charge of scalar field $S$.
New Yukawa and VL mass terms are then generated in the Lagrangian which, in the basis where the SM fields are diagonal, 
can be written as
\be\label{LagrLMLT}
\mathcal{L}\supset -\lam_{Q,i}S D_R\, d_{L,i} - \lam_{Q,k} (V^{\dag}_{\textrm{CKM}})_{ki}\,S\, U_R u_{L,i}
- m_{Q} \left(D_R D_L + U_R U_L\right) +\textrm{H.c.} \,,
\ee
with $u_{L,i}$, $d_{L,i}$ being the SM left-chiral quarks and $V_{\textrm{CKM}}$ the CKM matrix.
A sum over repeated indices 
$k,i=1,2,3$, labeling the quark generations, is implied. 

As is shown in greater detail in Appendix~\ref{app:frot}, 
after the U(1)$_X$ symmetry is spontaneously broken by the vev 
of the scalar field, $v_S$, the Lagrangian~(\ref{LagrLMLT}) is rotated to the physical basis, 
leading to flavor non-diagonal interactions of the $Z'$ boson with the SM quarks. One finds
\bea
g_L^{sb}&\approx & \pm g_X Q_S \frac{\sqrt{2}\, m_Q\, \lam_{Q,2}\, \lam_{Q,3}\, v_S^2}{\left(2 m_Q^2+\lam_{Q,2}^2\, v_S^2\right)  \sqrt{2\, m_Q^2+\left(\lam_{Q,2}^2+\lam_{Q,3}^2\right)v_S^2}}\,,\label{eq:delbsL}\\
g_R^{sb}&\approx & 0\,.\label{eq:delbsR}
\eea
A coupling $g_R^{sb}\neq 0$ can be easily generated by adding an extra VL spinor 
pair with the quantum numbers of the SM down-type quarks.

\subsection{Direct lepton couplings: $L_{\mu}-L_{\tau}$ symmetry\label{sec:mutau}}

The flavor-conserving effective couplings of $Z'$ to the muon, $g^{\mu\mu}_{L}$ and $g^{\mu\mu}_{R}$, 
can be generated either directly, or by the mixing of the muon with extra VL leptons. We will discuss the first
possibility in this subsection and the second in \refsec{sec:lepts}.

Direct $Z'$ couplings to the muons emerge if the SM is charged under U(1)$_X$. We employ the well known
$L_{\mu}-L_{\tau}$ gauge symmetry\cite{Foot:1990mn,He:1990pn,He:1991qd,Altmannshofer:2014cfa} 
to provide an anomaly-free 
SM extension generating the desired couplings. The SM lepton doublets, $l_{i=1,2,3}$, and singlets, $e_R$, $\mu_R$,
$\tau_R$, carry the following quantum numbers: 
\bea
l_1:(\mathbf{1},\mathbf{2},-1/2,0)\;\;\; & \quad & e_R:(\mathbf{1},\mathbf{1},1,0)\label{electrons}\\
l_2:(\mathbf{1},\mathbf{2},-1/2,1)\;\;\; & \quad & \mu_R:(\mathbf{1},\mathbf{1},1,-1)\label{muons}\\
l_3:(\mathbf{1},\mathbf{2},-1/2,-1) & \quad & \,\tau_R:(\mathbf{1},\mathbf{1},1,1)\label{taus}\,.
\eea
Since the $Z'$ interaction with the muon is VL in U(1)$_X$ one gets, straightforwardly,
\be\label{eq:delmu1}
g_V^{\mu\mu}=g_X\quad\quad\quad g_A^{\mu\mu}=0\,.
\ee

One can insert Eqs.~(\ref{eq:delbsL}) and (\ref{eq:delmu1}) into \refeq{Wils_generic} 
to obtain the WET Wilson coefficients. By recalling that $m_{Z'}^2= g_X^2 Q_S^2 v_S^2$ one gets
\bea\label{eq:C9_M1}
C_9^{\mu}\approx \mp \frac{1}{Q_S}\,\frac{2\Lambda_v^2}{V_{tb}V_{ts}^{\ast}}\, \frac{\sqrt{2}\, m_Q\, \lam_{Q,2}\, \lam_{Q,3}}{\left(2 m_Q^2+\lam_{Q,2}^2\, v_S^2\right)  \sqrt{2\, m_Q^2+\left(\lam_{Q,2}^2+\lam_{Q,3}^2\right)v_S^2}}\,, &  \quad & C_{10}^{\mu}=0\,.
\eea

In this case, the bound of \refeq{eq:bound1} applies. 
Assuming, for example, $Q_S=1$ and 
$|\lam|\equiv |\lam_{Q,i}|$, one gets 
\be
14\,\textrm{TeV} \lesssim m_Q/|\lam| \lesssim 22\,\textrm{TeV}
\ee
for $m_Q/|\lam| \approx m_{Z'}/g_X$. One gets instead
\be
20\,\textrm{TeV}\lesssim m_Q/|\lam| \lesssim 32\,\textrm{TeV}
\ee
for $m_{Z'}/g_X\ll m_Q/|\lam|$. Taken at face value, these solutions may be difficult to constrain directly at the LHC.

\subsection{Couplings through mixing: VL leptons with a U(1)$_X$ charge\label{sec:lepts}}

Alternatively, the effective couplings $g^{\mu\mu}_{L}$ and $g^{\mu\mu}_{R}$ can be generated in the same way as the quark couplings $g^{sb}_{L,R}$ in \refsec{sec:quark}. Unlike in \refsec{sec:mutau} we now keep the 
SM leptons uncharged under U(1)$_X$ and instead introduce a VL pair of left-chiral Weyl spinor doublets transforming 
under the total gauge group as
\be\label{eq:VLleptons}
L:(\mathbf{1},\mathbf{2},-1/2,Q_L) \qquad L':(\mathbf{1},\mathbf{\bar{2}},1/2,-Q_L)\,.
\ee
For $Q_L=Q_S$ ($Q_L=-Q_S$) the particle content allows for new Yukawa and mass terms
\be\label{eq:lep_yuk}
\mathcal{L}\supset \lam_{L,i} S^{(\ast)} L' l_i  + m_L L'L  +\textrm{H.c.}\,,
\ee
where $l_{i=1,2,3}$ are the SM lepton doublets (uncharged under U(1)$_X$) and, again, a 
sum over SU(2) and repeated flavor indices is implied.

Once the U(1)$_X$ symmetry is broken, the effective muon couplings take the form
\bea
g_L^{\mu\mu}&\approx &g_X\,Q_L\, \frac{\lam_{L,2}^2 v_S^2}{2 m_L^2+\lam_{L,2}^2 v_S^2}\,,\label{eq:gLmu}\\
g_R^{\mu\mu}&\approx &0\,,
\eea
which lead to 
\be\label{eq:C9_M2}
C_9^{\mu}=-C_{10}^{\mu}\approx \mp \frac{Q_L}{Q_S}\frac{\Lambda_v^2}{V_{tb}V_{ts}^{\ast}}\left[ \frac{\sqrt{2}\, m_Q\, \lam_{Q,2}\, \lam_{Q,3}}{\left(2 m_Q^2+\lam_{Q,2}^2\, v_S^2\right)  \sqrt{2\, m_Q^2+\left(\lam_{Q,2}^2+\lam_{Q,3}^2\right)v_S^2}}\right]
\left(\frac{\lam_{L,2}^2 v_S^2}{2 m_L^2+\lam_{L,2}^2 v_S^2}\right).
\ee
In this case, the bound of \refeq{eq:bound2} applies. Assuming
for example, $|\lam|\equiv |\lam_{Q,i}|\approx |\lam_{L,i}|$, one gets, 
\be\label{eq:uncos_qlz}
8\,\textrm{TeV}\lesssim m_{Q,L}/|\lam|
\lesssim 12\,\textrm{TeV}
\ee
for $m_{Q,L}/|\lam|\approx m_{Z'}/g_X$.
In the limit $m_{Z'}/g_X\approx m_L/|\lam| \ll m_Q/|\lam|$,
\be\label{eq:uncos_q}
11\,\textrm{TeV}\lesssim m_Q/|\lam| \lesssim 17\,\textrm{TeV}\,,
\ee
while typical scales for $m_{Z'}/g_X$, $m_L/|\lam|$ are expected to be in the 1-to-few TeV range. 
\bigskip

We conclude this section by pointing out that we have introduced in Eqs.~(\ref{eq:VLquarks}), (\ref{muons}),
(\ref{taus}), and (\ref{eq:VLleptons}) fermions that are charged under both the abelian U(1)$_Y$ and U(1)$_X$ gauge groups. 
As a direct consequence, abelian kinetic mixing will be generated in the Lagrangian and the $Z'$ will mix at the tree level with the neutral gauge bosons of the SM, as reviewed briefly in Appendix~\ref{app:vrot}. The size of the kinetic mixing in turn determines 
how sensitive the models described in this section
end up being to existing LHC searches for $Z'$ production with leptons in the final states.  

Finally, the rotations of the scalar fields from the gauge to the physical mass basis are presented in Appendix~\ref{app:srot}.

\section{Trans-Planckian boundary conditions }\label{sec:fpan}

\subsection{General notions\label{sec:as}}

The framework of AS we adopt in this work is based on the assumption that the SM and the NP particles introduced in \refsec{sec:models} couple above the Planck scale, at energy $k\geq M_{\textrm{Pl}}=10^{19}\gev$, to quantum gravity or some other NP responsible for generating a fixed point 
in the trans-Planckian RG flow of the beta functions of all dimensionless couplings. The beta functions of the SM and NP receive in the trans-Planckian regime parametric corrections, 
\bea\label{eq:gauge-Yuk}
\beta_g&=&\beta_g^{\textrm{SM+NP}}-g\,f_g, \nonumber \\
\beta_y&=&\beta_y^{\textrm{SM+NP}}-y\,f_y, \nonumber \\
\beta_\lam&=&\beta_\lam^{\textrm{SM+NP}}-\lam\,f_\lam, 
\eea
where $\beta_x\equiv dx/d\log k$, and $g$, $y$, $\lam$ are the set of all gauge, Yukawa and scalar couplings, respectively. The effects of new trans-Planckian interactions are parameterized with fixed effective couplings $f_g$, $f_y$, and $f_\lam$. In agreement with theoretical computations in asymptotically safe gravity, the latter are assumed to be universal, in the sense that they differ according to the type of matter interaction (gauge, Yukawa, scalar) but do not change with the quantum numbers of the particles involved.

The parameters $f_g$, $f_y$, and $f_\lam$ should be eventually determined from the gravitational dynamics\cite{Zanusso:2009bs,Daum:2009dn,Daum:2010bc,Folkerts:2011jz,Oda:2015sma,Eichhorn:2016esv,Eichhorn:2017eht,Christiansen:2017cxa}.
In the framework of quantum gravity, calculations utilizing the functional RG have shown that $f_g$ is likely to be non-negative, irrespective of the chosen RG scheme\cite{Folkerts:2011jz}, and that $f_g>0$ is required to make the gauge sector asymptotically free. 
On the other hand, the status of the leading-order gravitational correction to the Yukawa couplings, $f_y$, remains to some extent more uncertain. 
A number of simplified models were analyzed in the literature\cite{Rodigast:2009zj,Zanusso:2009bs,Oda:2015sma,Eichhorn:2016esv}, 
but no definite conclusions regarding the size and sign of $f_y$ are available to our knowledge. The same level of ignorance applies to the gravity contribution to the scalar coupling beta function, $f_\lam$\cite{Wetterich:2016uxm,Eichhorn:2017als,Pawlowski:2018ixd,Wetterich:2019rsn,Eichhorn:2020kca,Eichhorn:2020sbo,Eichhorn:2021tsx}.

One ought to keep in mind that the theoretical determination of the parameters $f_g$, $f_y$, and $f_\lam$ is marred by large uncertainties, which are related to the choice of truncation of the gravitational action and, within a chosen truncation, the cutoff-scheme dependence\cite{Reuter:2001ag,Narain:2009qa}. Depending on the number of operators considered in the effective action\cite{Lauscher:2002sq,Codello:2007bd,Benedetti:2009rx,Falls:2017lst,Falls:2018ylp} various determinations can differ by up to 50-60\%\cite{Dona:2013qba}. To bypass these obstacles, we follow in this work 
the effective approach adopted in some recent articles\cite{Eichhorn:2017ylw,Eichhorn:2018whv,Reichert:2019car,Alkofer:2020vtb,Kowalska:2020gie,Kowalska:2020zve,Kowalska:2022ypk}, in which $f_g$, $f_y$, and $f_\lam$ are treated as free parameters to be determined by matching the SM couplings onto their low-energy measurements. 
The ensuing fixed-point analysis of the matter beta functions defines in turn a set of boundary conditions for the NP couplings which, unlike the SM ones, have not been directly measured yet.

A fixed point of the system defined in \refeq{eq:gauge-Yuk} is given by any set $\{g^\ast,y^\ast,\lam^\ast\}$, generically denoted with an asterisk,
for which $\beta_g(g^\ast,y^\ast,\lam^{\ast})=\beta_y(g^\ast,y^\ast,\lam^{\ast})=\beta_{\lam}(g^\ast,y^\ast,\lam^{\ast})=0$. The structure of the fixed point is then
determined by linearizing the system of RGEs. Let us define $\{\alpha_i\}\equiv\{g,y,\lam\}$; one thus computes in the vicinity of the fixed point the stability matrix $M_{ij}$\,:
\be\label{stab}
M_{ij}=\partial\beta_i/\partial\alpha_j|_{\{\alpha^{\ast}_i\}}\,,
\ee
whose eigenvalues are $-\theta_i$, with $\theta_i$ being the critical exponents. If $\theta_i>0$ the corresponding eigendirection is UV-attractive and dubbed as {\it relevant}. All RG trajectories along this direction asymptotically reach the fixed point, thus a deviation of a relevant coupling from the fixed point introduces a free parameter in the theory. One can then use this freedom to adjust the coupling at some high trans-Planckian scale to match an eventual measurement in the IR.  If $\theta_i<0$ is negative, the corresponding eigendirection is UV-repulsive and dubbed as {\it irrelevant}. In such a case only one trajectory exists that the coupling's flow can follow towards the IR, thus providing potentially a specific prediction for its value at the experimentally accessible scale. Finally, $\theta_i=0$ corresponds to a \textit{marginal} eigendirection characterized by the RG flow that is logarithmically slow and one needs to go beyond the linear approximation to determine whether a fixed point is attractive or repulsive.

\subsection{Fixed-point analysis of the $Z'$ models}

\subsubsection{Gauge-Yukawa system\label{sec:gayusys}}

The one-loop RGEs of the gauge and Yukawa couplings for the models defined in \refsec{sec:models} are presented in Appendix~\ref{app:RGEsKM} and Appendix~\ref{app:RGEs}.
We work in the basis where the quark mass matrix is diagonal, 
since it facilitates the matching of the quark Yukawa couplings onto the experimentally measured values. 
Consequently, we take into account the running of the elements of the CKM matrix\cite{Babu:1987im,Sasaki:1986jv,Barger:1992pk,Kielanowski:2008wm}.
Since we are mainly interested in the parameters affecting $b\to s$ transitions, 
we restrict the analysis of the quark Yukawa RGE system to two generations, the second and the third, in the ``down-origin'' approach (see Appendix~\ref{app:RGEs}). The CKM matrix is thus approximated by an orthogonal $2\times 2$ matrix, parameterized by one single rotation angle.

The scalar potential couplings do not affect the running of the gauge-Yukawa system at one loop and therefore they can be treated separately. We will come back to discussing the fixed-point structure of the scalar potential in \refsec{sec:scal_FP}.

The minimal set of couplings pertinent to the analysis of the gauge-Yukawa system is the following:
\bea
\textrm{SM}:&& g_3\,,\; g_2\,,\; g_Y\,,\;y_t\,,\;y_b\,,\;V_{33}\,,\label{eq:paramsSM}\\
\textrm{NP:} && g_D,\;g_\eps,\;\lambda_{Q,2},\;\lambda_{Q,3},\;\lambda_{L,2}\,,\label{eq:paramsBSM}
\eea
where $g_3$, $g_2$, and $g_Y$ are the gauge couplings of SU(3)$_\textrm{c}$, SU(2)$_L$, and U(1)$_Y$, respectively, $y_t$ and $y_b$ indicate the Yukawa couplings of the top and bottom quarks, and $V_{33}$ belongs to the CKM matrix. The running NP gauge couplings $g_D$ and $g_{\epsilon}$ are related to the gauge coupling $g_X$ and abelian gauge boson kinetic mixing $\epsilon$ as defined in 
Appendix~\ref{app:RGEsKM}. The Yukawa couplings of the quarks of the first two generations and of the leptons are not included in the system, as their impact on the running of the parameters defined in \refeq{eq:paramsSM} and \refeq{eq:paramsBSM} is negligible. We associate all of them with relevant directions of a Gaussian fixed point in the trans-Planckian UV\cite{Alkofer:2020vtb} and so their IR values can always be reached.

Let us start the fixed-point analysis by focusing on the gauge sector. In what follows, we will indicate the fixed-point values of dimensionless couplings  with an asterisk. Given the expectation that $f_g>0$, the non-abelian gauge couplings develop non-interactive (or Gaussian) fixed points,
\be
g_3^\ast=0,\qquad g_2^\ast=0\,,
\ee
as required by the low-energy phenomenology. Both of them correspond to relevant directions in the coupling space. 

\begin{table}[t]
\footnotesize
\begin{center}
\begin{tabular}{|c|c|c|c|}
\hline
Model & Sec. reference & Fermion charges & $\widetilde{Q}_{YX}$ \\
\hline
1A & \refsec{sec:lepts} & $Q_{\textrm{SM}}=0$, $Q_L=Q_S$ & 0 \\
\hline
1B & \refsec{sec:lepts} & $Q_{\textrm{SM}}=0$, $Q_L=-Q_S$ & $8/3\, Q_S$ \\
\hline
2 & \refsec{sec:mutau} & $Q_{\mu}=1$, $Q_{\tau}=-1$ & $4/3\, Q_S$ \\
\hline
\end{tabular}
\caption{The effective charge $\widetilde{Q}_{YX}$, introduced in \refeq{eq:qu1x} of Appendix~\ref{app:RGEsKM}, determines the fixed-point value of kinetic mixing for the three scenarios considered in this study.}
\label{tab:mod_kin}
\end{center}
\end{table}

The abelian sector is described by the three couplings $g_Y$, $g_D$, and $g_\eps$. 
Following the AS ansatz, we assume that quantum gravitational interactions 
are going to tame their pathological UV behavior. The system thus 
develops a fully interactive fixed point, 
\be\label{eq:gyast}
g_Y^\ast=4\pi\sqrt{\frac{f_g}{\widetilde{Q}_Y}}\,,\quad g_D^\ast=4\pi\sqrt{\frac{f_g \widetilde{Q}_Y}{\widetilde{Q}_Y \widetilde{Q}_X-\widetilde{Q}_{YX}^2}}\,,\quad  g_\eps^\ast=-4\pi \widetilde{Q}_{YX}\sqrt{\frac{f_g }{\widetilde{Q}_Y^2 \widetilde{Q}_X-\widetilde{Q}_Y\widetilde{Q}_{YX}^2}}\,,
\ee
corresponding to irrelevant directions in the coupling space. 
Incidentally, a fixed point with $g_D^\ast=0$ is also allowed, but it is not phenomenologically interesting as it would imply a non-interacting $Z'$. The quantities $\widetilde{Q}_{Y}$, $\widetilde{Q}_{X}$, and $\widetilde{Q}_{YX}$ are defined in 
Eqs.~\eqref{eq:qu1}--\eqref{eq:qu1x} in Appendix~\ref{app:RGEsKM}. It is possible to make the kinetic mixing vanish at the fixed point, 
provided $\widetilde{Q}_{YX}=0$. One may use \refeq{eq:qu1x}
to determine whether or not this happens in
the models introduced in \refsec{sec:models}. We analyze three cases, summarized in \reftable{tab:mod_kin}:
\begin{itemize}
    \item Model~1A, with VL leptons introduced in \refeq{eq:VLleptons} of \refsec{sec:lepts}
    and $Q_L=Q_S$
    \item Model~1B, same as above but $Q_L=-Q_S$
    \item Model~2, with $L_{\mu}-L_{\tau}$ symmetry as described in \refsec{sec:mutau}.
\end{itemize}
In Model~1A the kinetic mixing $g_{\epsilon}$ vanishes at the fixed point. Moreover, it
is not generated through the running, as the corresponding beta function is multiplicative in $g_\eps$, see \refeq{eq:be_comp} in Appendix~\ref{app:RGEsKM}. 

Note that the irrelevant (predictive) nature of trans-Planckian fixed points for the abelian gauge couplings is essential to be able to uniquely determine the value of the effective parameter $f_g$. This is done by 
matching $g_Y$ onto its phenomenological value in the IR, 
$g_Y(M_t)=0.358$. One obtains $f_g=0.012$ in models 1A, 1B, and $f_g=0.010$ in Model~2. 

The second quantum gravity parameter, $f_y$, is fixed by postulating 
that at least one of the SM Yukawa couplings presents a UV interactive fixed point\cite{Eichhorn:2018whv}, 
\be\label{eq:yuk_match}
y_t^\ast=F_t\left(f_g,f_y\right)\qquad \textrm{and/or}\qquad y_b^\ast=F_b\left(f_g,f_y\right)\,.
\ee
The flow along an irrelevant direction, from the fixed point down to the IR, 
is then matched onto the value of the corresponding quark mass. Finally, we associate the CKM matrix element $V_{33}$ with a relevant direction and assume it is zero at the fixed point\cite{Alkofer:2020vtb}:
\be
V_{33}^\ast= 0\,.
\ee

Once the $f_g$, $f_y$ parameters are extracted by matching the SM couplings to their IR value, one proceeds to determine the UV fixed points of the NP Yukawa couplings in \refeq{eq:paramsBSM}. The process will eventually lead to specific low-scale predictions, as long as those couplings correspond to irrelevant directions in theory space. 
Out of the full set of fixed points, we extract those with maximal predictivity, which belong to two classes of trans-Planckian solutions,
\bea\label{eq:YukFP}
&\textrm{FP}_{M,a}:& \lambda_{Q,2}^{\ast}\neq 0,\;\lambda_{Q,3}^{\ast}= 0\,,\nonumber \\
&\textrm{FP}_{M,b}:& \lambda_{Q,2}^{\ast}= 0,\;\lambda_{Q,3}^{\ast}\neq 0\,,
\eea
where the subscript $M=$\,1A, 1B, 2 spans the models in \reftable{tab:mod_kin}. Moreover, in models 1A, 1B we require
\be
\lambda_{L,2}^\ast\neq 0\,.
\ee
Note that a fully interactive solution with \textit{real} 
Yukawa couplings $\lambda_{Q,2}^{\ast} \neq 
\lambda_{Q,3}^{\ast}\neq 0$ does not exist. Conversely, we neglect the solution  
$\lambda_{Q,2}^\ast=\lambda_{Q,3}^\ast=0$ along a relevant direction because that choice would not lead to an enhancement in predictivity with respect to the simplified model approach. 

\begin{table}[t]
\footnotesize
\begin{center}
\begin{tabular}{|c|cc|ccc|cccc|}
\hline
 &$f_g$ & $f_y$ & $g_Y^\ast$ & $g_D^\ast$ & $g_\eps^\ast$ & $y_t^\ast$ & $\lambda_{Q,3}^\ast$ & $\lambda_{Q,2}^\ast$ & $\lambda_{L,2}^\ast$ \\
\hline
$\textrm{FP}_{\textrm{1A},a}$ & 0.012 & 0.0025 & 0.498 & 0.418 & 0 & 0.406 & 0 & 0.072 & 0.648 \\
$\textrm{FP}_{\textrm{1A},b}$ & 0.012 &  0.0029 & 0.498 & 0.418 & 0 & 0.424 & 0.200 & 0 & 0.610 \\
\hline
$\textrm{FP}_{\textrm{1B},a}$ & 0.012 &  0.0026  & 0.498 & 0.436 &  0.151 & 0.417 & 0  & 0.163 & 0.586 \\
$\textrm{FP}_{\textrm{1B},b}$ & 0.012 &  0.0034 & 0.498 & 0.436 &  0.151 & 0.452 & 0.264 & 0 & 0.547 \\
\hline
$\textrm{FP}_{2,a}$ & 0.010 &  0.0018 & 0.479 & 0.366 & 0.069 & 0.428 & 0 & 0.354 & -- \\
$\textrm{FP}_{2,b}$ & 0.010 & 0.0037 & 0.479& 0.366 & 0.069 & 0.453 & 0.379& 0 & --\\
\hline
\end{tabular}
\caption{Trans-Planckian fixed-point values for the gauge and Yukawa couplings 
and the corresponding gravity parameters $f_g$ and $f_y$ for the 
scenarios investigated in this work.}
\label{tab:FPall}
\end{center}
\end{table}

The list of all fixed points of phenomenological interest, together 
with the values assumed by $f_g$ and $f_y$ for the three scenarios considered in this study 
is presented in \reftable{tab:FPall}.\footnote{The fact that $f_g$ and $f_y$ assume sizable numerical values insures that the predictions for the NP Yukawa couplings change minimally under the addition of perturbative 2-loop contributions to the beta functions. This variation is far smaller than the experimental uncertainty on the Wilson coefficients of the effective Hamiltonian and, as a consequence, it does not affect the predicted range of the mass parameters.} We shall henceforth set $Q_S=-1$. The predictions of Model~1 do not depend on the explicit value of $Q_S$, as everything rescales with the product $g_D Q_S$, which is a constant of the RG flow. That is not the case for Model~2, where some sensitivity to the value of $Q_S$ arises due to the asymmetry between the U(1)$_X$ charge assigned to the scalar field $S$ 
and the $L_{\mu}-L_{\tau}$ charge of the leptons. The dependence of the resulting phenomenology on the specific choice of $|Q_S|\neq 1$ may be non-trivial and would require a case-by-case study.
  
In all cases, we can only reproduce the IR 
value of the bottom mass if $y_b^{\ast}=0$ (relevant). 
Conversely, the top Yukawa coupling at the fixed point is irrelevant and thus in principle the parameter $f_y$ should be determined by the top quark mass like in \refeq{eq:yuk_match}. However, there exists a minimum $f_y$ allowing $y_b^{\ast}=0$ to be relevant and therefore we adopt that as the $f_y$ value of choice. 

\begin{table}[b]
\footnotesize
\begin{center}
\begin{tabular}{|c|ccc|cccc|}
\hline
 & $g_Y(k_0)$ & $g_D(k_0)$ & $g_\eps(k_0)$ & $y_t(k_0)$ & $\lambda_{Q,3}(k_0)$ & $\lambda_{Q,2}(k_0)$ & $\lambda_{L,2}(k_0)$ \\
\hline
$\textrm{FP}_{\textrm{1A},a}$  & 0.364 & 0.305 & 0 & 1.08 & -0.381 & 0.016 & 0.823 \\
$\textrm{FP}_{\textrm{1A},b}$  & 0.364 & 0.305 & 0 & 1.09 & 0.034 & 0.803 & 0.606 \\
\hline
$\textrm{FP}_{\textrm{1B},a}$ & 0.363 & 0.318 &  0.110 & 1.05  & -0.612 & 0.296 & 0.652 \\
$\textrm{FP}_{\textrm{1B},b}$ & 0.363 & 0.318 &  0.110 & 1.08 & 0.004 & 0.874 & 0.499 \\
\hline
$\textrm{FP}_{2,a}$ & 0.363 & 0.277 & 0.052 & 1.05 & -0.905 & 0.298 &-- \\
$\textrm{FP}_{2,b}$ & 0.363 & 0.277 & 0.052 & 1.10 & 0.040 & 0.988 & --\\
\hline
\end{tabular}
\caption{Values of the gauge and Yukawa couplings of the scenarios investigated in this work at the reference ``collider'' scale of $k_0=2\tev$.}
\label{tab:Predall}
\end{center}
\end{table}

Once the quantum gravity contribution to the RGEs 
is turned off below $M_{\textrm{Pl}}$, the system of the couplings is run down to an 
indicative ``collider'' scale, which we set at $k_0=2\,\textrm{TeV}$. 
For the models considered in this work, the values of 
the couplings at $k_0$ are presented in \reftable{tab:Predall}. Note that in the two-family approximation adopted here we do not generate radiatively $\lam_{Q,1}$. However, one should bear in mind that in the more realistic three-family approach a small $\lam_{Q,1}(k_0)\lesssim 10^{-5}$ 
can arise due to the additive contributions in its RGE,
\be
\frac{d \lam_{Q,1}}{dt}\simeq -\frac{y_t^2\, V_{31}}{16\,\pi^2}\left(V_{32} \lam_{Q,2}+V_{33} \lam_{Q,3} \right)\,.
\ee

An attentive look at the low-scale predictions in \reftable{tab:Predall} shows that the running kinetic mixing coupling 
$g_{\epsilon}$ vanishes at all scales in Model~1A but is instead different from zero in models~1B and 2, in agreement with the corresponding 
value of $\widetilde{Q}_{YX}$ reported in \reftable{tab:mod_kin}. One can thus compute with \refeq{eq:epsilon} in Appendix~\ref{app:RGEsKM}  the corresponding
$\epsilon(k_0)=0.29$ in Model~1B, and $\epsilon(k_0)=0.14$ in Model~2.
As will be discussed in detail in \refsec{sec:constr}, 
these values lead to the direct exclusion of Model~1B and Model~2 at the LHC.
For this reason, we shall focus entirely on Model~1A for the remainder of this subsection.

 \begin{figure}[t]
	\centering%
	    \subfloat[]{%
		\includegraphics[width=0.46\textwidth]{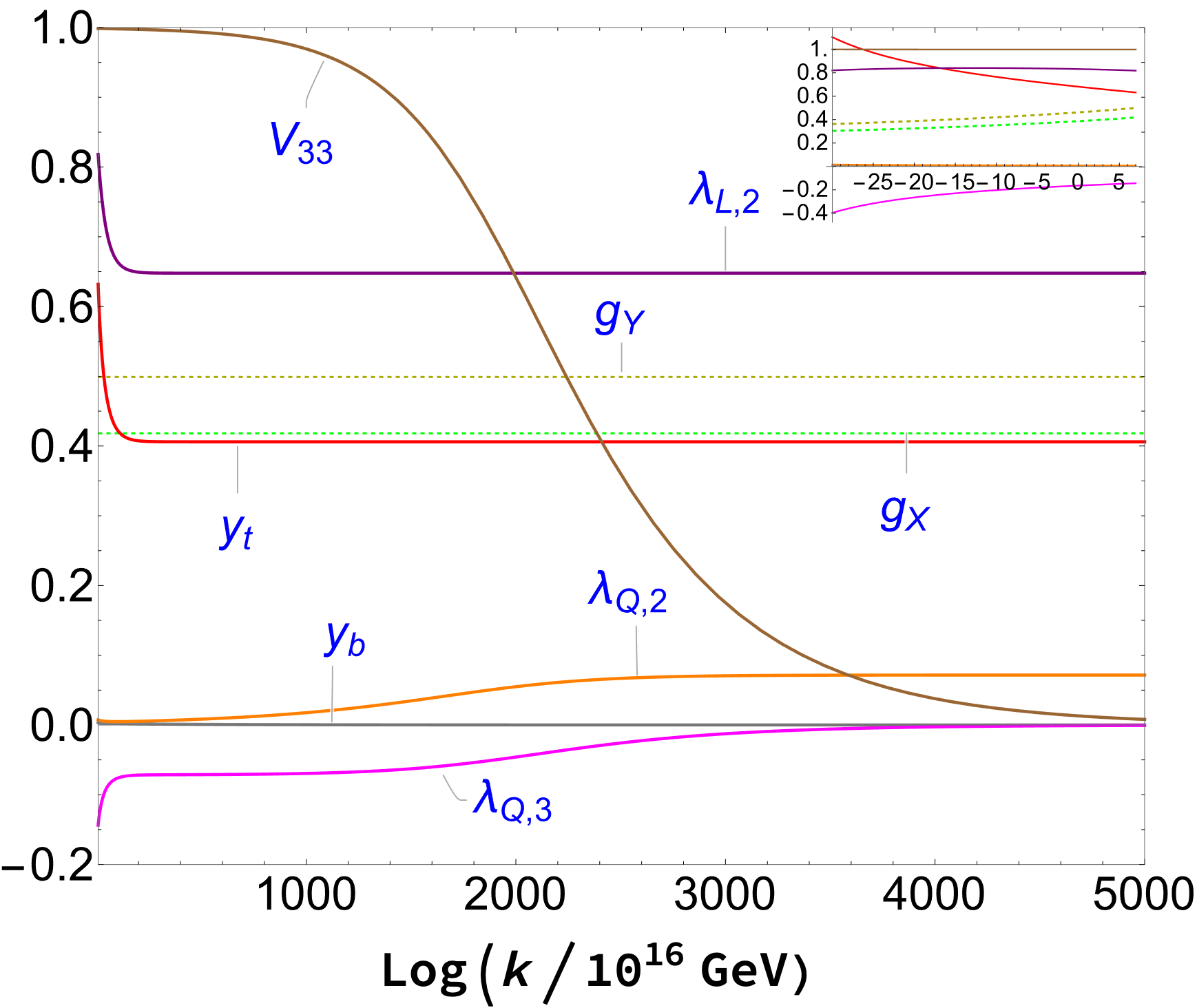}}
		\hspace{0.8cm}
		\subfloat[]{%
		\includegraphics[width=0.46\textwidth]{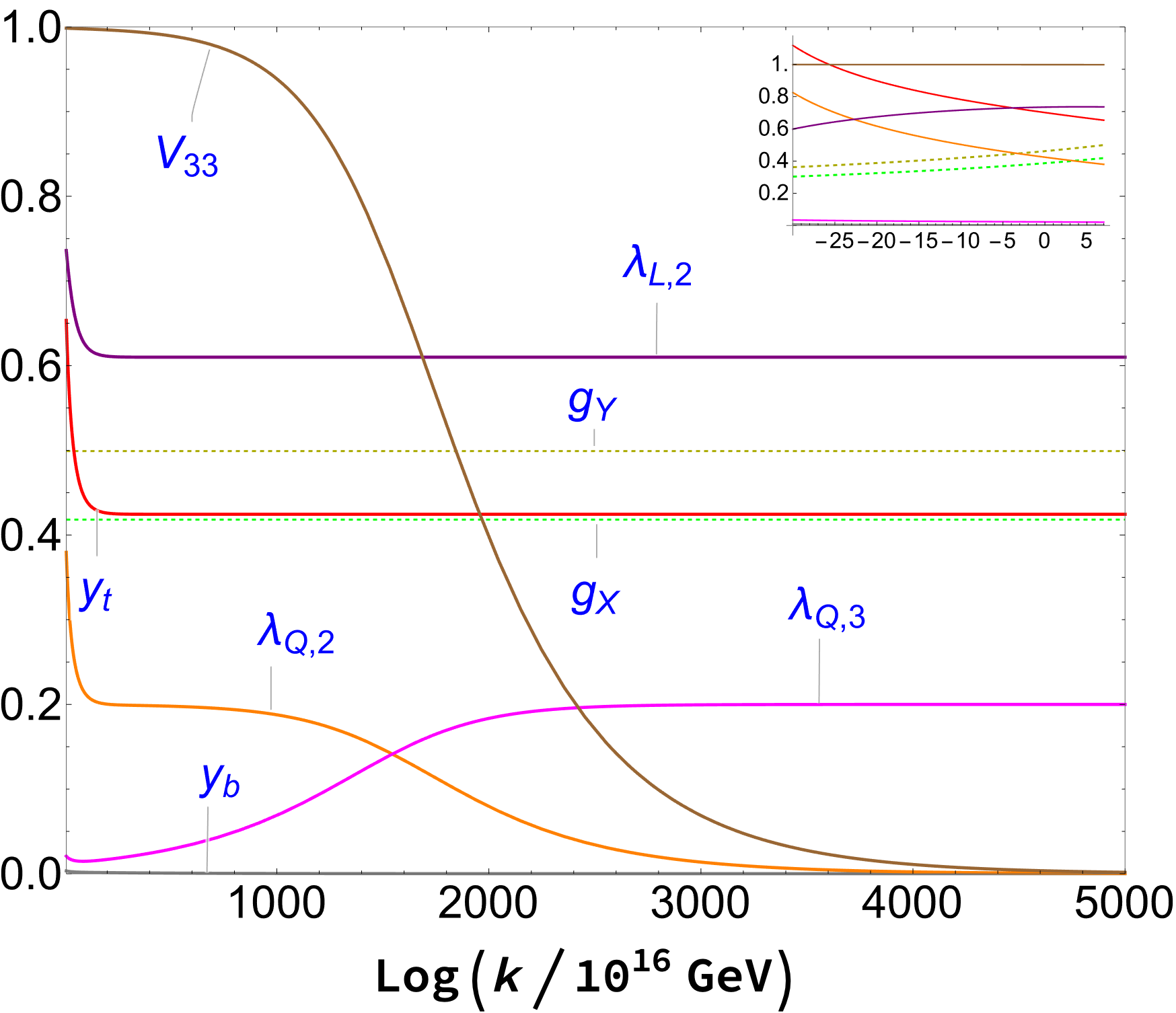}}
\caption{(a) RG flow of the gauge and Yukawa couplings from trans-Planckian energies 
down to the EWSB scale for a scenario characterized by UV fixed point FP$_{\textrm{1A},a}$. 
The sub-Planckian flow is shown in the inset panel. (b) The same for FP$_{\textrm{1A},b}$. }
\label{fig:flow}
\end{figure}

We show the trans-Planckian flow from FP$_{\textrm{1A},a}$ in \reffig{fig:flow}(a) and the flow from   
FP$_{\textrm{1A},b}$ in \reffig{fig:flow}(b). At the fixed point FP$_{\textrm{1A},a}$ 
coupling $\lam_{Q,2}$ is of the irrelevant type. Conversely,
the system of couplings ($V_{33}$, $\lam_{Q,3}$) spans a 
2-dimensional submanifold in theory space, 
on which both couplings are relevant but do 
not correspond to eigendirections of the stability matrix. 
An attentive look at \reffig{fig:flow}(a)
shows the likely appearance, on the left-hand side of the plot,
where the flow approaches from above the Planck scale, 
of an additional IR-attractive fixed point for the RGE trajectories, distinct from FP$_{\textrm{1A},a}$. 
This fixed point, confirmed by the numerical analysis, is characterized by $V_{33}^{\ast}=1$, $\lam_{Q,3}^{\ast}=-0.1$, $\lam_{Q,2}^{\ast}=0$, $y_b^{\ast}=0$, all of them irrelevant. An IR-attractive fixed point of similar features was already observed for the leptoquark case in Ref.\cite{Kowalska:2020gie};
it effectively washes out much of the freedom associated with relevant directions in the Yukawa-coupling theory space and sets 
the typical value the couplings of the system take at the Planck scale. 

Note also that the exact mass of the top quark cannot be exactly matched at the EWSB scale. Because there is a minimum $f_y$ value required to make $y_b^\ast=0$ relevant, the running $y_t$ always leads to the top 
quark being heavier than the measurement by about 10\%. This is unfortunate, but it seems to be a feature appearing in other studies 
of trans-Planckian boundary conditions, in the SM and NP alike, see, \textit{e.g.}, Refs.\cite{Alkofer:2020vtb,Kowalska:2020gie}. 
Possibly, a detailed analysis of the 
RGEs at higher loop order may shed some light on this issue. 
However, since a very precise determination of the top mass is not essential to the extraction of the predictions of these $Z'$ models for flavor and collider physics, we will leave the investigation of this topic to future work.

At the fixed point of type FP$_{\textrm{1A},b}$ both the NP Yukawa couplings of the quark sector span irrelevant directions. While $\lam_{Q,3}$ corresponds to an eigenvector of the stability matrix, the flow of $\lam_{Q,2}$ close to the fixed point is entirely dictated by the UV hypercritical surface relating it with the (relevant) CKM matrix element $V_{33}$,  $\lam_{Q,2}(t)= \mathcal{F}(V_{33}(t))$.
The trans-Planckian flow of the parameters of the system is presented in \reffig{fig:flow}(b). As was the case for FP$_{\textrm{1A},a}$, the trajectories eventually cross over to the basin of attraction of an IR-attractive 
fixed point characterized by $V_{33}^\ast=1$, $\lam_{Q,2}^{\ast}=0.2$, $\lam_{Q,3}^{\ast}= 0$, $y_b^\ast = 0$. 
Note, however, that in neither Model~$\textrm{1A},a$ nor Model~$\textrm{1A},b$
the new IR-attractive fixed point is ever fully reached, as the requirement of reproducing the low-energy value of the CKM matrix element $V_{33}$ implies that the gravitational effects decouple before the system stabilizes. As was the case in Model~$\textrm{1A},a$, the experimentally favored value of the top Yukawa coupling is overshot in Model~$\textrm{1A},b$ by about 20\%.

\subsubsection{Scalar potential\label{sec:scal_FP}}

The scalar potential of the models considered in this work is given by
\be\label{eq:sca_pot}
V\left(\left|h\right|^2,\left|S\right|^2\right)=-\mu_h^2 h^{\dag}h+\lam_h\left(h^{\dag}h\right)^2+\mu_S^2\, S^{\dag}S
+\lam_S \left(S^{\dag}S\right)^2+\lam_{hS}\left(S^{\dag}S\right)\left(h^{\dag}h\right),
\ee
where $\mu_h^2$ and $\lam_h$ are the mass parameter and the quartic coupling of the SM Higgs boson doublet $h$, $\mu_S^2$ and $\lam_S$ are the mass parameter and the quartic coupling of the scalar $S$,
and $\lam_{hS}$ is the portal coupling. The 1-loop RGEs of the dimensionless parameters of the potential are given in Appendices \ref{app:RGEs_model1}-\ref{app:RGEs_mutau}.

As was mentioned above, 
the quartic couplings of the scalar potential do not enter at one loop the RGEs 
of the gauge-Yukawa system and as such their fixed-point properties  
will not impact the flavor phenomenology in a significant way. 
On the other hand, predictions can be derived for the mass of the heavy scalar $m_{H_2}$, 
where $H_2$ is a mixture of 
the SM Higgs and the scalar $S$ (dominated by the latter, see Appendix~\ref{app:srot}).

The qualitative properties of the fixed-point $(\lam_h^\ast,\,\lam_S^\ast,\,\lam_{hS}^\ast)$ 
are the same in all the considered scenarios. The sign of the 
real fixed points and corresponding critical exponents are presented in \reftable{tab:SCA_FP} for $f_{\lam}<0$.
In three out of the four points the portal coupling develops a quasi-Gaussian fixed point, 
$|\lambda_{hS}^\ast|\ll 10^{-2}$. In all four points $\lambda_{hS}^\ast$ 
is positive and IR attractive for $f_\lam<0$ and negative and UV attractive for $f_\lam>0$.

\setlength\tabcolsep{0.25cm}
\begin{table}[t]\footnotesize
\begin{center}
\begin{tabular}{|c|ccc|ccc|}
\hline
 & $\lam_h^\ast$ & $\lam_S^\ast$ & $\lam_{hS}^\ast$ & $\theta_h$ & $\theta_S$ & $\theta_{hS}$ \\
\hline
FP$_{\textrm {A}}$ & $>0$ & $>0$ & $\approx 0^+$ & $-$ & $-$  & $-$ \\
FP$_{\textrm {B}}$ & $>0$ & $<0$ & $\approx 0^+$ & $-$ & $+$ & $-$ \\
FP$_{\textrm {C}}$ & $<0$ & $>0$ & $\approx 0^+$ & $+$ & $-$   & $-$\\
FP$_{\textrm {D}}$ & $<0$ & $<0$ & $>0$ & $+$ & $+$   & $-$ \\
\hline
\end{tabular}
\caption{The sign of real fixed points of the system $(\lam_h,\,\lam_S,\,\lam_{hS})$ (left-hand side box) and 
the signs of the corresponding critical exponents (right-hand side box) for $f_{\lam}<0$.}
\label{tab:SCA_FP}
\end{center}
\end{table}

\begin{figure}[t]
	\centering%
	    \subfloat[]{%
		\includegraphics[width=0.45\textwidth]{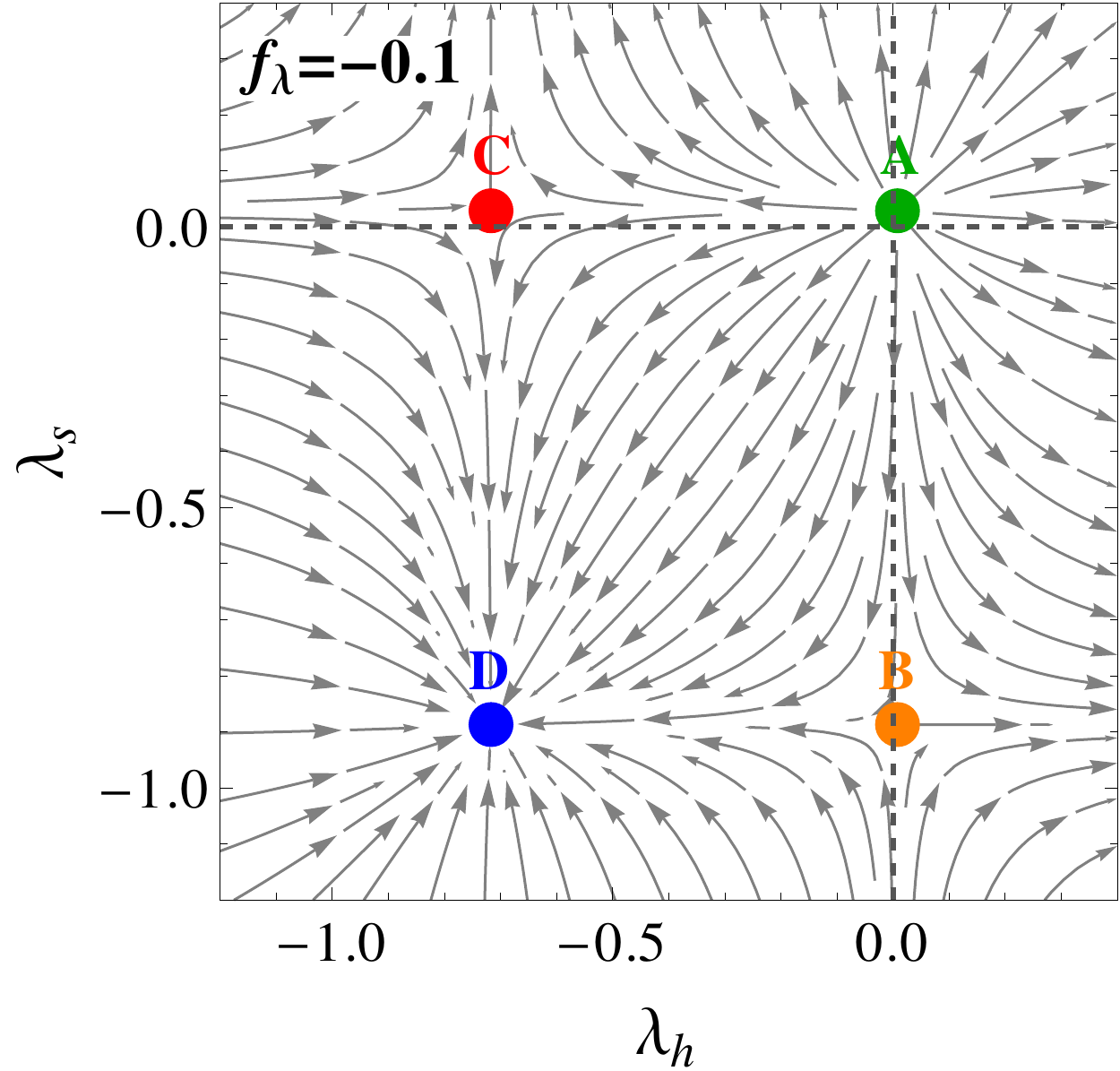}}
		\hspace{0.8cm}
		\subfloat[]{%
		\includegraphics[width=0.47\textwidth]{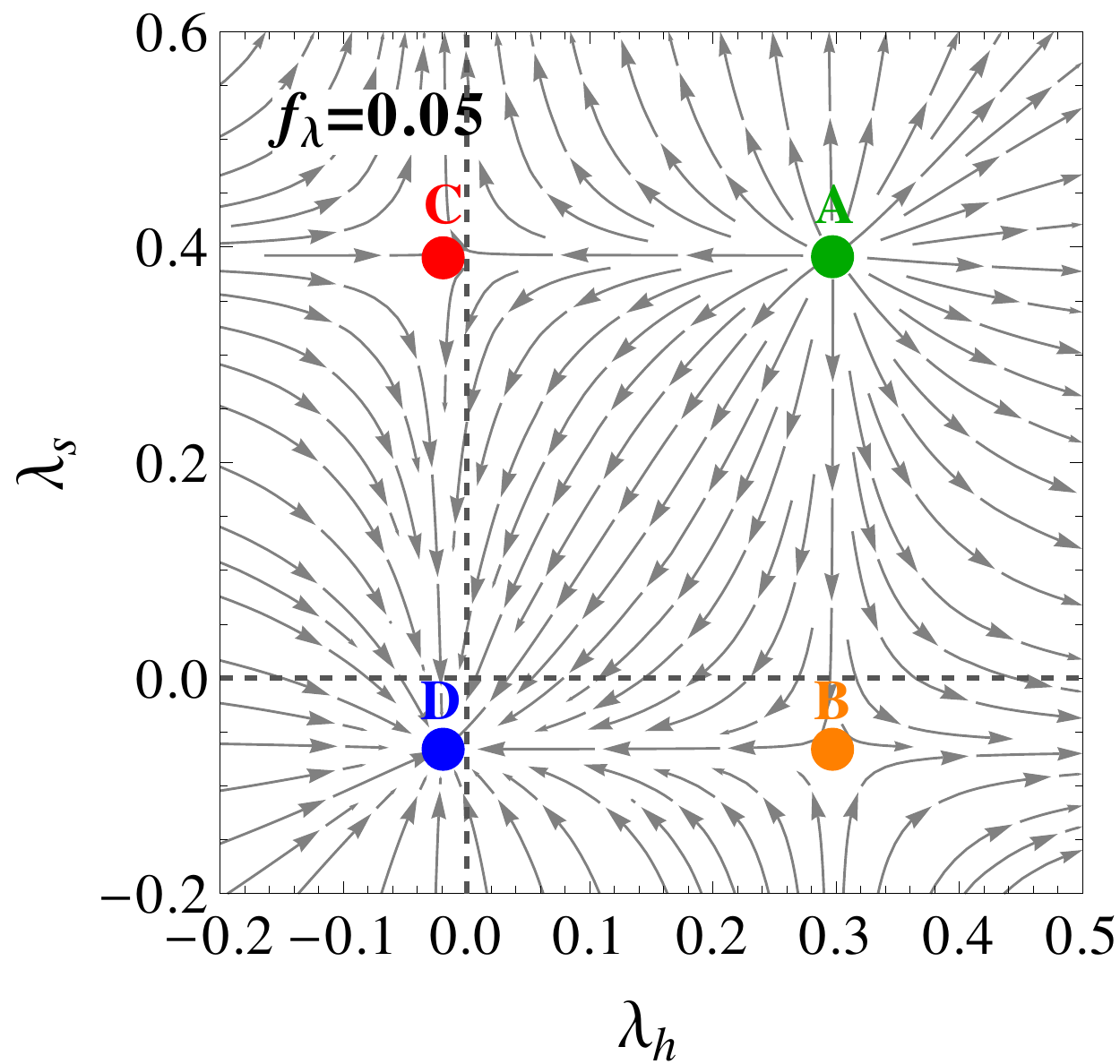}}
\caption{(a) Phase space diagram in the $(\lam_h,\lam_S)$ plane for $f_\lam=-0.1$ and the couplings
set at  FP$_{\textrm{1A},a}$ in \reftable{tab:FPall}. 
(b) Phase space diagram for $f_\lam=0.05$. The RG flow directions point towards the UV. Color dots correspond to the fixed points listed in \reftable{tab:SCA_FP}.}
\label{fig:PhaseDiag_pos}
\end{figure}

The actual value of the couplings depends on the model at hand and, most importantly, 
on the size and sign of $f_{\lam}$, which should eventually emerge from a quantum gravity calculation.
The dependence of the system on $f_{\lam}$ is shown in
\reffig{fig:PhaseDiag_pos}. In both panels the fixed points 
of \reftable{tab:SCA_FP} are marked with color dots. The arrows always point 
towards the UV. 

In \reffig{fig:PhaseDiag_pos}(a) the phase diagram is shown for a negative value $f_{\lam}=-0.1$, which is large compared to the typical size of the terms involving 
the gauge and Yukawa couplings in the RGEs of Appendix~\ref{app:RGEs}
(recall that $f_g,\,f_y \lesssim 10^{-2}$, see \reftable{tab:FPall}). 
It forces the appearance of a pseudo-Gaussian fixed point,
FP$_{\textrm {A}}$, fully irrelevant, which makes the potential stable at the Planck scale.
Infrared-attractive fixed points of the scalar potential are often discussed in the literature for their predictivity\cite{Shaposhnikov:2009pv,Eichhorn:2017als,Pawlowski:2018ixd}. However, 
since in these models we overshoot the top mass value at low energies -- see discussion above -- 
we also overshoot the Higgs mass at the EWSB scale. The behavior of $y_t$ and $\lam_h$ is strongly connected and does not depend much on the specific form of the Higgs quartic-coupling RGE which, if anything, features in these models additional positive terms with respect to the SM, allowing for an easier running. As was the case for the top mass determination, a full two-loop analysis may be required to shed more light on this issue.

As one increases $f_{\lam}$, towards less negative values, to zero and then positive values, 
the four fixed points move up and to the right with respect to \reffig{fig:PhaseDiag_pos}(a). 
In \reffig{fig:PhaseDiag_pos}(b) we show the case of positive $f_\lam=0.05$, which is also 
relatively large with respect to the typical size of the RGE terms of 
the gauge and Yukawa couplings. In this case, one can refer to the fully relevant fixed point 
FP$_{\textrm {D}}$, which implies that the three quartic couplings become effectively  
free parameters and issues with the IR value of the Higgs mass do not arise. 
The scalar potential, on the other hand, 
may possibly become unstable at the Planck scale or, in the best-case scenario, 
metastable in a fashion similar to the SM\cite{Degrassi:2012ry}. A detailed analysis of stability of the scalar potential exceeds the scope of this paper.   

We conclude, finally, by pointing out that $\lam_S^\ast$ assumes different values in the two panels in \reffig{fig:PhaseDiag_pos}, depending on the selected $f_{\lam}$. Nevertheless, a unique low-scale prediction for $\lam_S(M_t)$ (and therefore $m_{H_2}$) can be extracted, as the sub-Planckian flow of $\lam_S$ is predominantly determined by the top and NP Yukawa couplings and 
is quite insensitive to the exact value of 
$\lam_S$ at the Planck scale (as long as it remains small). In Model~$\textrm{1A},a$ we obtain the NP couplings
\be\label{eq:scal_pred}
\lam_S(173\gev)=0.18\,,\qquad \lam_{hS}(173\gev)=0.1\,,
\ee
which lead to a prediction for the heavy scalar mass
\be\label{eq:scalmass} 
m_{H_2}\approx 2.02\, m_{Z'}\,.
\ee

A non-zero value of the portal coupling $\lam_{hS}$ introduces mixing between 
the SM Higgs and the scalar singlet $S$. 
For the values of the scalar potential couplings given in \refeq{eq:scal_pred}, \refeq{eq:mixan} in
Appendix~\ref{app:srot} can be used to obtain
\be
\sin\alpha_H\approx\frac{20\gev}{m_{Z'}}.
\ee
This is far below the experimental upper bound which reads $\sin\alpha_H<0.2$\cite{Robens:2021rkl}.

\subsubsection{Comment on the stability of the fixed points under extra NP \label{sec:stabil}}

Our assumption of a ``desert''  of new particles, from collider energies up to the Planck scale and above, seems to be at odds with the need of explaining other observational realities -- like the existence of dark matter and the smallness of neutrino masses -- that are not directly addressed in the $Z'$ models introduced in this work. As a matter of fact, any construction in which dark matter and/or a mechanism for neutrino masses involved new interactions with order-one couplings to the SM would likely lead to the appearance of sizable terms in the RGEs of Appendix~\ref{app:RGEs}. As these potentially induce a shift in the position of the trans-Planckian fixed points, such uncertainty might lead some readers to question the accuracy of the predictions we derive in the next section.

On the other hand, the impact of additional NP can be minimized or canceled altogether if one assumes that the SM extension addressing phenomena beyond the flavor anomalies mostly comprises feebly interacting particles. We assume that this is always the case in this work, and we construct in Appendix~\ref{app:feeble_ext} 
a specific, quantitative example of a viable possibility in this framework, bearing in mind that a detailed phenomenological analysis of the dark matter and neutrino sectors exceeds the scope of this work. 

\section{Experimental constraints on the $Z'$ models}\label{sec:constr}

Since the new fermions introduced in \refsec{sec:models} are VL 
they neither generate gauge anomalies nor induce via loops new 
axial-vector contributions to the decay of the SM gauge bosons, which are strongly constrained. Moreover, since the U(1)$_X$ symmetry is broken by the vev of a SM singlet, we do not expect large contributions to the oblique parameters. In other words, the minimal models we have chosen are safe from precision-physics constraints.

The trans-Planckian fixed-point analysis of the gauge-Yukawa system allows one to predict the low-scale values of the irrelevant couplings. Since the RG flow of the NP Yukawa couplings is very slow over the phenomenologically interesting $\sim$TeV range, these can effectively be treated as constants at the low scale, which leaves the VL fermion masses $m_Q$, $m_L$, and the $Z'$ mass $m_{Z'}$ as the only remaining free parameters. This is in agreement with the fact that Lagrangian mass parameters are canonically relevant in AS. 

The parameter space of the $Z'$ models is subject to several constraints. These include: 
$m_{Z'}$-dependent bounds on the kinetic mixing of neutral gauge bosons; 
flavor bounds on $B_s$-meson mixing; neutrino trident production; and
direct bounds on the mass of the VL fermions and $Z'$ boson from NP searches at the LHC. 

\subsection{Kinetic mixing}\label{sec:kinmix}

The kinetic mixing of heavy $Z'$ ($m_{Z'}>m_Z$)
can be experimentally tested by looking for narrow $Z'$ resonances decaying to muons and/or electrons\cite{Jaeckel:2012yz}.
The dominant production channels proceed through the couplings to the first quark generation which, in the limit $m_Q, m_L\gg m_{Z'}$, are given in \refeq{eq:SM_coup} of Appendix~\ref{app:vrot}.
The most recent measurements by ATLAS\cite{ATLAS:2019erb} and CMS\cite{CMS:2021ctt}, based on 140\invfb\ data set in proton–proton collisions at the centre-of-mass energy of $\sqrt{s}=13\tev$ provide upper bounds on the $p\,p\to Z'\to l^+\,l^-$ cross-section as a function of the $Z'$ boson mass. These in turn can be translated onto the limit on the kinetic mixing \eps.

\begin{figure*}[t]
	\centering%
		\includegraphics[width=0.7\textwidth]{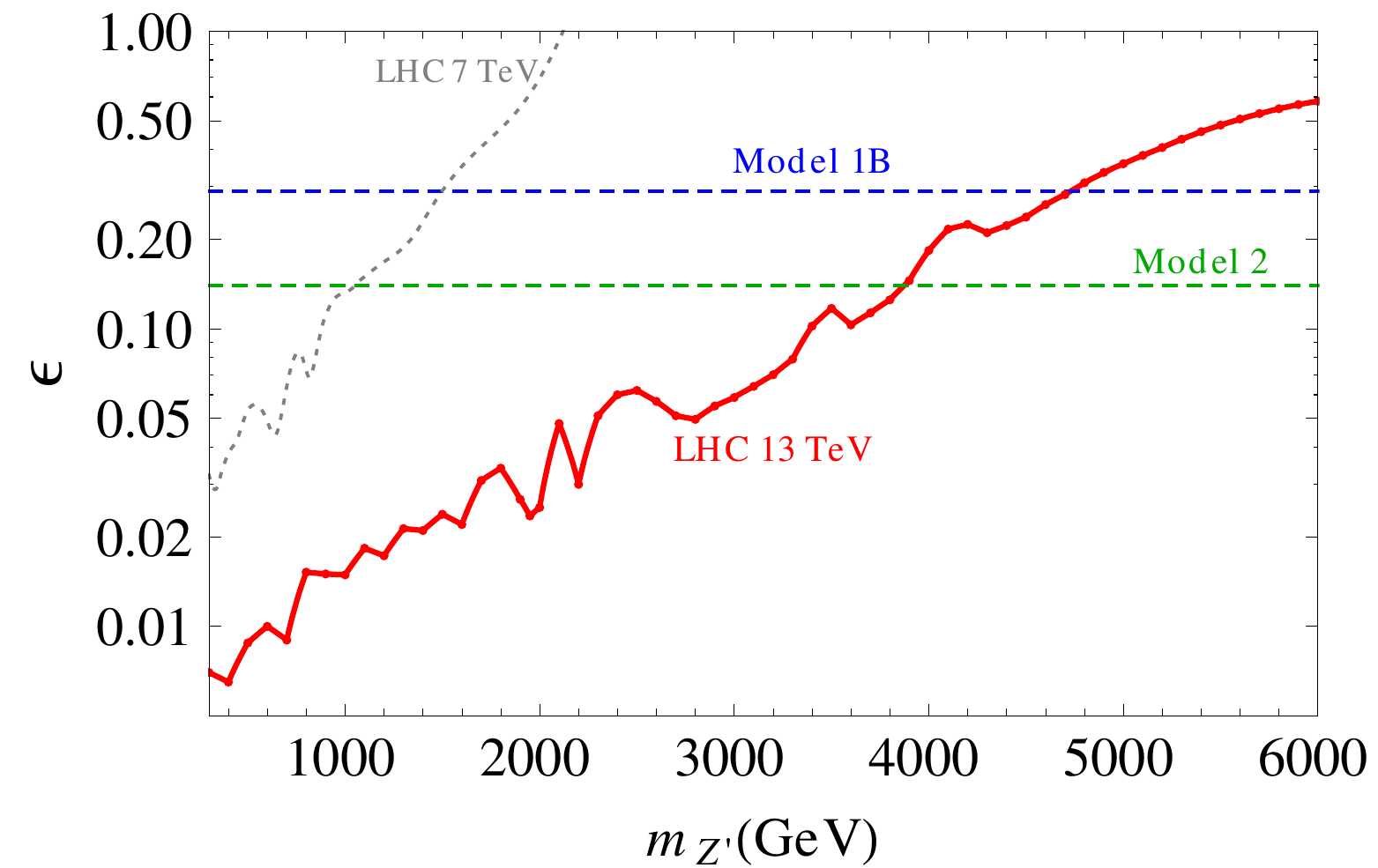}
\caption{Our computation of the experimental bound on the kinetic mixing parameter $\eps$ as a function of the $Z'$ mass. In solid red our results based on the 13\tev\ ATLAS analysis\cite{ATLAS:2019erb} are shown. In dashed gray we show for comparison the 7~TeV exclusion bound derived in Ref.\cite{Jaeckel:2012yz}. 
Dashed blue and green line indicate the predictions of Models 1B and 2, respectively.}
\label{fig:KinMix13}
\end{figure*}

We compute for the first time the lower bound on $m_{Z'}$, 
based on the ATLAS 13~TeV search. It is presented 
in \reffig{fig:KinMix13} as a red solid line. The cross section for the process $p\,p\to Z'\to l^+\,l^-$, where $l=e,\,\mu$, is simulated with \madgr\ v3.4.1\cite{Alwall:2014hca}. 
Note that our computation is completely model-independent.
For comparison, we show as a gray dotted line the 
7~TeV exclusion bound derived in Ref.\cite{Jaeckel:2012yz}.  The predictions for the kinetic mixing parameter in Model~1B and Model~2, which are computed with \refeq{eq:epsilon} in Appendix~\ref{app:RGEsKM}, are superimposed in the plot as
a dashed blue~(1B) and a dashed green line~(2). 

\subsection{$B$-meson mixing and neutrino trident production}

$B_s - \bar{B}_s$ transitions can constrain directly the mass parameters.
We use\cite{Zyla:2020pdg} 
\begin{equation}\label{eq:bsbound}
    R_{\Delta M_s}=\frac{\Delta M_s^{\textrm{exp.}}}{\Delta M_s^{\textrm{SM}}}-1=-0.09\pm 0.08
\end{equation}
to impose a bound on $m_Q$, $m_{Z'}$. 
The quantity of interest is parameterized as\cite{Altmannshofer:2014rta}
\be\label{bsmix}
R_{\Delta M_s}=\frac{v_h^2 \left(g_L^{sb}\right)^2}{m_{Z'}^2}
\left[\frac{g_2^2}{16\pi^2}\left(V_{tb}V_{ts}^{\ast}\right)^2 S_0\right]^{-1},
\ee
where $v_h=246\,\textrm{GeV}$ is the SM Higgs vev, 
$g_2$ is the SU(2)$_L$ gauge coupling, $S_0 \approx 2.3$ is a loop function, and 
$g_L^{sb}$ is given in \refeq{eq:delbsL}.
\bigskip

If $Z'$ couples to muon neutrinos, a strong enhancement in the neutrino trident production from scattering on atomic nuclei, $N$:
$N \nu \rightarrow \nu N \mu^+ \mu^-$, is expected\cite{Altmannshofer:2014pba}. The corresponding cross section has been measured by the CCFR\cite{CCFR:1991lpl} and CHARM-II\cite{CHARM-II:1990dvf} collaborations in agreement with the SM prediction. For $m_{Z'}> 1\,\textrm{GeV}$, these results translate into the universal lower bound on the new gauge boson mass,
\begin{equation}\label{eq:trident}
\left(\frac{m_{Z'}}{g_{\nu}}\right)>500\gev,
\end{equation}
where $g_{\nu}$ denotes a generic coupling of $Z'$ to neutrinos. In the case of the $L_\mu-L_\tau$ symmetry (Model~2) one finds $g_\nu=g_X$, while in the scenario where the interactions of $Z'$ with the SM lepton doublet arise through the mixing with VL leptons (Model~1) one finds $g_\nu= g_L^{\mu\mu}$, with $g_L^{\mu\mu}$ given in \refeq{eq:gLmu}.  
\bigskip

 \begin{figure}[t]
	\centering%
	    \subfloat[]{%
		\includegraphics[width=0.45\textwidth]{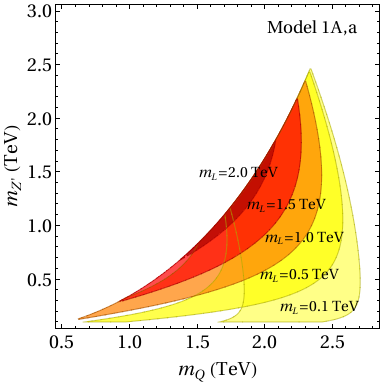}}
		\hspace{0.8cm}
		\subfloat[]{%
		\includegraphics[width=0.45\textwidth]{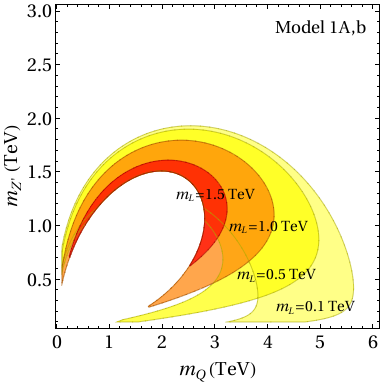}}
\caption{(a) The parameter space of Model~$\textrm{1A},a$  
allowed by the flavor anomalies and the constraints from $B_s$-meson mixing and neutrino trident production in the $(m_Q,m_{Z'})$ plane. In different colors the parameter space corresponding to selected choices of $m_L$. 
(b) Same for Model~$\textrm{1A},b$.}
\label{fig:ParSpace1A}
\end{figure}

We present in \reffig{fig:ParSpace1A}(a), 
in the ($m_Q$, $m_{Z'}$) plane for different choices of $m_L$, 
the parameter space of Model~$\textrm{1A},a$ 
allowed by a combination of the flavor anomalies constraint given in \refeq{eq:bound2}, 
the bound from $B_s$-meson mixing in \refeq{eq:bsbound}, and the neutrino trident production limit in \refeq{eq:trident}. In \reffig{fig:ParSpace1A}(b) we show the corresponding parameter space for Model~$\textrm{1A},b$.  Note that there exists an upper bound on the $Z'$ mass, $m_{Z'}\lsim 2.5\tev$, which is known to be mostly due to the $B_s$-mixing constraint. 

The VL fermion mass parameters are bounded as well: $0.5\tev \lesssim m_Q\lesssim 2.7\tev$ in Model~$\textrm{1A},a$ and $m_Q\lsim 5.5\tev$ in Model~$\textrm{1A},b$;  $m_L\lsim 2.3\tev$ in Model~$\textrm{1A},a$ and $m_L\lsim 1.6\tev$ in Model~$\textrm{1A},b$.
Compare these determinations to the mostly unconstrained parameter space of 
the simplified model, Eqs.~(\ref{eq:uncos_qlz}) and (\ref{eq:uncos_q}):
the strong predictions for the models' couplings that arise from AS open up 
the enticing possibility of testing the parameter space entirely in current and future collider searches. We investigate such  possibility in \refsec{sec:collider}. 

By again applying Eqs.~(\ref{eq:bound1}), (\ref{eq:bound2}), and (\ref{eq:bsbound}) with the coupling values of \reftable{tab:Predall} one can easily check that a $2\sigma$ upper 
bound on the $Z'$ mass, $m_{Z'}\lesssim 2.6\tev$, exists in models 1B and 2 as well.  
On the other hand, we read in \reffig{fig:KinMix13} that the 95\%~C.L. 
lower bound on the $Z'$ mass due to the kinetic mixing reads
\bea
\textrm{Model 1B:} & \quad &  m_{Z'}\gsim 4.7\,\textrm{TeV}\,,\label{eq:kin_1B}\\
\textrm{Model 2:} & \quad &  m_{Z'}\gsim 3.9\,\textrm{TeV}\,,\label{eq:kin_2}
\eea
which unequivocally lead to the full exclusion of Model~1B and Model~2 in the AS framework. 
Incidentally, the same conclusion does not apply to Model~1A, as the parameter $g_{\eps}$ (and thus $\epsilon$) remains identically zero in its run from the Planck scale to the scale of decoupling of the largest VL mass, $\max(m_Q,m_L)$.
Below that scale kinetic mixing is loop-generated, 
proportionally to the product of hypercharge $Y_i$ and U(1)$_X$ charge $Q_i$ of the fermions of label $i$ transforming under both symmetries. The final result,
\be
\epsilon \approx \cos\theta_W \frac{g_Y g_X}{12\pi^2} \sum_{i\,\in\, \textrm{fermions}}
Y_i Q_i \ln \frac{m_Q^2}{m_L^2}\approx 10^{-3}\,,
\ee
remains however too small for Model~1A to be sensitive to the bound in \reffig{fig:KinMix13}.

\subsection{Collider searches\label{sec:collider}}

The Lagrangians of \refeq{LagrLMLT} and \refeq{eq:lep_yuk} imply the presence of NP particles beyond the SM. The theory contains 
an up and a down quark of the fourth generation, $U_4$ and $D_4$, degenerate at the tree level, 
a charged lepton 
and Dirac heavy neutrino, $E_4$ and $N_4$, again degenerate, 
as well as a heavy $Z'$ gauge boson and a heavy scalar $H_2$. These particles can in principle be directly produced at the LHC. In this subsection we briefly discuss the possibility of constraining the parameter space of Model~$\textrm{1A}$ with the results of the LHC Run~II, as well as some prospects for the future upgrades.

\paragraph{$Z'$ boson} Even in the absence of kinetic mixing, the quark fusion process $u(s)\,u(s) \to Z'\to \mu^+\mu^-$ can impose constraints on the $Z'$ mass, as was observed, \textit{e.g.}, in Ref.\cite{Darme:2018hqg}. 
Since quark fusion can only proceed through the mixing of VL quarks 
with the SM quarks $u$ and $s$, the corresponding cross section strongly depends on the size of the 
NP Yukawa couplings $\lam_{Q,2}$ and $\lam_{Q,3}$. 

Following the strategy described in \refsec{sec:kinmix}, we apply to the process $u(s)\,u(s) \to Z'\to \mu^+\mu^-$ the  experimental bounds on the cross section from the ATLAS\cite{ATLAS:2019erb} and CMS\cite{CMS:2021ctt} 140\invfb\ narrow resonance searches. In Model~$\textrm{1A},b$ this translates into a 95\%~C.L. lower bound on the $Z'$ mass,
\be\label{eq:LHCstr}
m_{Z'}\gsim 5\tev\,.
\ee
The bound is quite strong, the main reason being the large value of the $Z'$ coupling to 
up and strange quark pairs due to large $\lam_{Q,2}(k_0)=0.803$ (see \reftable{tab:Predall}) and 
$V_{us}\, \lam_{Q,2}(k_0)\approx 0.18$.

As \reffig{fig:ParSpace1A}(b) shows, the constraints from the flavor anomalies and 
$B_s$ mixing yield
\be\label{eq:bfrdelm2}
m_{Z'}\lesssim 2\,\textrm{TeV}\,,
\ee
which makes Model~$\textrm{1A},b$ excluded. Only Model~$\textrm{1A},a$ remains untested by the narrow resonance searches as the production cross section is in this case suppressed by four orders of magnitude due to the small value $\lam_{Q,2}(k_0)=0.016$. 

\paragraph{VL fermions}
The VL leptons can be pair produced at the LHC via Drell-Yan processes. 
Their physical mass at the tree level reads
\be
m_{E_4,N_4}=\sqrt{m^2_L+\lam_{L,2}^2 m^2_{Z'}/2g^2_X}\,.
\ee
In Model~$\textrm{1A},a$
this means that $m_{E_4,N_4}>m_{Z'}$ since, 
given the values in \reftable{tab:Predall}, $\lam^2_{L,2}/2 g_X^2>1$.

On the other hand, the hierarchy between $E_4$ and the scalar $H_2$ depends on the relative size of the Lagrangian parameter $m_L$ with respect to $m_{Z'}$ (recall from \refsec{sec:scal_FP} that $m_{Z'}\approx 0.5\, m_{H_2}$). If $m_L \lesssim 0.6\,m_{Z'}$, the scalar is heavier than the VL lepton and the dominant decay channel is $E_4\to Z'\mu$, followed by $Z'\to \mu^+\mu^-$ and $Z'\to \nu_\mu \nu_\mu$ with a  branching ratio of $50\%$ each. These decay chains lead to a characteristic multilepton plus missing energy (MET) signature, reminiscent of that arising in the production of supersymmetric charginos. Conversely, if $m_L>0.6\,m_{Z'}$, 
the decay channel $E_4\to H_2\, \mu$ becomes kinematically allowed, with characteristic leptons and jets in the final state. 

As for the VL quarks of the fourth generation, their physical mass is given at the tree level by
\be
m_{U_4,D_4}\approx \sqrt{m^2_Q+\left(\lam_{Q,2}^2+\lam_{Q,3}^2 \right) m^2_{Z'}/2g^2_X}\,.
\ee
They are predominantly produced at the LHC via strong interactions. As long as $m_{U_4,D_4}\gg m_{Z'}$ the dominant decay channels are $U_4(D_4)\to Z'\,t(b)$, $U_4\to H_2\,t(b)$, resulting in a multijets plus leptons signature which can be tested by the LHC searches tailored to look for supersymmetric gluinos decaying through stops. 

 \begin{figure}[t]
	\centering%
		    \subfloat[]{%
		\includegraphics[width=0.42\textwidth]{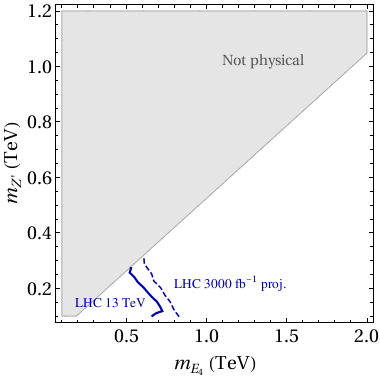}}
				    \subfloat[]{%
		\includegraphics[width=0.42\textwidth]{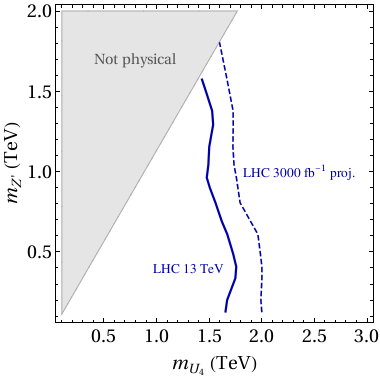}}\\
				    \subfloat[]{%
		\includegraphics[width=0.47\textwidth]{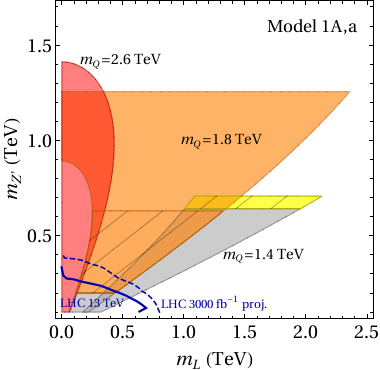}}
\caption{(a) 95\%~\cl\ lower exclusion bound on the mass of heavy VL leptons $E_4$ and $N_4$ from the CMS 13 TeV search\cite{CMS:2017moi} (solid blue line) and the projection for the future luminosity of $3000$ fb$^{-1}$ (dashed blue line). 
(b) 95\%~\cl\ lower exclusion bound on the mass of heavy VL quarks $U_4$ and $D_4$ from the 13 TeV ATLAS searches\cite{ATLAS:2019fag} and\cite{ATLAS:2021twp} (solid blue line) and the corresponding projection (dashed blue line).
(c) In the $(m_L,m_{Z'})$ plane, the parameter space of Model~$\textrm{1A},a$ allowed by the flavor anomalies and the constraints from $B_s$-meson mixing and neutrino trident production. The present 95\%~\cl\ exclusion bound on the VL quark mass is denoted in solid gray, while the corresponding projection for $3000$ fb$^{-1}$ with dashed shadings. The 95\%~\cl\ exclusion bound on the VL lepton mass and corresponding projection are indicated by a solid blue line and a dashed blue line, respectively.}
\label{fig:ParSpace1A_LHC}
\end{figure}

To impose the LHC bounds on the complex parameter space of Model~$\textrm{1A},a$
we use \chmt\ (latest version directly from its \texttt{github} master branch)\cite{Read:2002hq,Cacciari:2005hq,Cacciari:2008gp,Cacciari:2011ma,deFavereau:2013fsa}.
Events are generated using \texttt{Herwig}~v7.2.3\cite{Bellm:2015jjp,Bellm:2019zci} thanks to a \texttt{SARAH} generated \texttt{UFO} model file\cite{Degrande:2011ua,Staub:2012pb,Staub:2013tta}.
Parameter points are passed to \texttt{Herwig} via an SLHA\cite{Skands:2003cj,Allanach:2008qq} output generated from a \texttt{SARAH} created \texttt{SPheno}-like\cite{Porod:2003um,Porod:2011nf} spectrum generator.\footnote{In the scan we employ the \texttt{pyslha} package\cite{Buckley:2013jua} to handle \texttt{SPheno} input files.}
We simulate production of all two-particle combinations from the list $\{E_4, \bar{E}_4, U_4, \bar{U}_4, D_4, \bar{D}_4, N_4, \bar{N}_4\}$ (where some of the combinations are identically 0 due to various conservation laws).
In \reffig{fig:ParSpace1A_LHC}(a) we present as a solid dark blue line the recast of the 95\%~\cl\ lower exclusion bound on the mass of heavy VL leptons $E_4$ and $N_4$ from the CMS 13\tev\ analysis\cite{CMS:2017moi}.  The exclusion is slightly stronger than the one presented in Fig.~19 of Ref.\cite{CMS:2017moi} for a simplified model of Higgsino production because the branching ratio of $Z'\to \mu^+ \mu^-$, down in the decay chain of $E_4$, is larger than the corresponding branching ratio of the SM $Z$ boson emitted by the Higgsino. 
We also show as a dashed blue line the projections for the future high-luminosity reach, which we obtained, very conservatively, by rescaling the signal, the background yield, and the background uncertainty 
by a factor $L/L_0$, where $L=3000\,\textrm{fb}^{-1}$ 
is the default higher luminosity and $L_0$ the current one, 
and setting the number of observed events equal to the background.  

In \reffig{fig:ParSpace1A_LHC}(b) we show the lower exclusion bound on the mass of the heavy VL quarks $U_4$ and $D_4$. Mainly two searches contribute to the exclusion bound: at $m_{Z'}< m_{U_4,D_4}$ the decay chain 
$U_4\to Z' t$ is tested by the ATLAS search\cite{ATLAS:2019fag}. At the upper edge, $m_{Z'}\approx m_{U_4,D_4}$, decays to the top quark are not kinematically allowed 
and the most aggressive bound is placed by the ATLAS search\cite{ATLAS:2021twp}. The dashed blue line shows, again, the reach obtained by the rescaling procedure described above. 

We summarize the constraints on the parameter space of Model~$\textrm{1A},a$ in
\reffig{fig:ParSpace1A_LHC}(c). The region allowed by the flavor anomalies and $B_s$-mixing is shown in the 
($m_L$, $m_{Z'}$) plane for different choices of the VL mass $m_Q$. All parameter space with $m_Q<1.3\tev$ is excluded by the ATLAS search\cite{ATLAS:2019fag}. For 
$m_Q> 1.3\tev$ the regions excluded by hadronic production searches are shaded in solid gray. Conversely, the current bound from Drell-Yan production of heavy VL leptons, which is dominated by the CMS 13\tev\ analysis\cite{CMS:2017moi}, is 
superimposed on the plot as a solid blue line, and the high-luminosity projections of \reffig{fig:ParSpace1A_LHC}(a)
correspond to the dashed blue line. 
We also cover with dashed shadings the regions corresponding to the high-luminosity projection 
of \reffig{fig:ParSpace1A_LHC}(b). We evince that future coverage of the hadronic production will exclude the entire parameter space up to approximately $m_Q\approx 1.5\tev$, and significantly test the regions at higher $m_Q$ values. The regions that remain beyond reach in hadronic production, on the other hand, like the area shaded in red corresponding to $m_Q=2.6\tev$, will be probed deeply by the Drell-Yan searches.  

\section{Summary and conclusions}\label{sec:summary}

In this paper, we considered two simple and well-known $Z'$ extensions of the SM providing a solution to the long-standing flavor anomalies in $b\to s \mu^{+}\mu^{-}$ transitions. We constrained them with a double ``chokehold'' consisting, in the deep UV, of well-motivated boundary conditions for the otherwise free parameters of the models and, in the IR, of a comprehensive study of the most recent experimental bounds applying to them. 

In the models we considered the extra U(1)$_X$ 
gauge symmetry is broken by the vev of a scalar SM singlet and the
flavor non-diagonal couplings of the $Z'$ to the $s$ and $b$ quarks are generated via the mixing with heavy quarks of a VL fourth generation, which carry U(1)$_X$ charge. While in Model~1 we apply the same mechanism to the lepton sector 
and generate the coupling of the $Z'$ to muons via the mixing of the latter with 
heavy VL leptons, in Model~2 the $L_{\mu}-L_{\tau}$ symmetry is employed to generate the $Z'\mu\mu$ coupling directly. 
Those two constructions lead to different expectations for the $Z'/\gamma$ 
kinetic mixing. Moreover, we considered two different choices of the charges within Model~1 itself, 
also leading to different expectations for the kinetic mixing, which we dub as Model~1A and Model~1B. 

In order to constrain the resulting three scenarios in the UV, we chose to apply 
the ansatz of trans-Planckian AS. This is based on 
the assumption that, while the particles at the collider scale should be responsible for the flavor anomalies, 
a ``desert'' devoid of any other NP state extends all the way up to the Planck scale.
At $M_{\textrm{Pl}}$, interactions of the matter states with quantum gravity or some other physics lead to the appearance of fixed points for the beta functions of the dimensionless couplings in the Lagrangian. We thus performed a fixed-point analysis of the three models to determine the relevant (UV-attractive) and irrelevant (IR-attractive) directions in theory space. While the former identify the effectively free parameters of the Lagrangian, the latter yield predictions for the couplings, which in turn can be tested experimentally once the gravity interactions are decoupled at $M_{\textrm{Pl}}$ and the system is run back down to the EWSB scale.

Thanks to the trans-Planckian fixed-point analysis we were able to derive a fairly precise determination
of the abelian kinetic mixing $\epsilon$, the NP Yukawa couplings, and the scalar quartic couplings. 
We then compared the obtained results with the Wilson coefficient ranges 
favored in the global EFT analyses to identify the allowed NP mass ranges. The emerging parameter space 
was finally subjected to a phenomenological analysis of the most recent ATLAS and CMS constrains, which we simulated numerically. 
In particular, we computed the most recent 95\%~C.L. bound on the ($m_{Z'}$, $\epsilon$) plane 
from $Z'$ production searches, based on the 13~TeV data set.
We found that this search excludes two out of the three scenarios (1B and 2) 
due to the insurmountable constraint on the large kinetic mixing generated along the flow from the Planck scale down. 

Conversely,  the same strong bound can be evaded in Model~1A with an appropriate choice of the U(1)$_X$ charges.  We have thus subjected this last scenario to the bounds from direct production of VL heavy quarks and leptons at the LHC. By using \texttt{CheckMate}, we identified the parameter space excluded at the 95\%~C.L., and we computed the projections for the planned increase in luminosity in future runs. We showed that direct NP searches are effective in excluding the parameter space consistent with the flavor anomalies and have the potential to bite into the unconstrained regions 
in even greater depth.

Overall, this work confirms the findings of several recent studies, from our group and others,  
pointing to the dramatic boost in predictivity that can be obtained by endowing NP scenarios with UV 
completions based on the well-motivated assumption of trans-Planckian AS. If the flavor anomalies are eventually 
confirmed to be real 
in future experimental data at LHCb and Belle~II, the possibility of observing NP particles directly at the LHC rather than 
in a more powerful, yet-to-be-conceived machine, may point to the existence of a UV completion 
based on the ultra-high scale properties of quantum gravity. 

\bigskip
\begin{center}
\textbf{ACKNOWLEDGMENTS}
\end{center}
\noindent 
AC, WK, and EMS are supported by the National Science Centre (Poland) under the research Grant No.~2020/38/E/ST2/00126. KK and DR are supported by the National Science Centre (Poland) under the research Grant No.~2017/26/E/ST2/00470. The use of the CIS computer cluster at the National Centre for Nuclear Research in Warsaw is gratefully
acknowledged.

\bigskip

\appendix

\addcontentsline{toc}{section}{Appendices}

\section{Rotations to the physical basis}

\subsection{Fermions\label{app:frot}}

We construct the mass matrix $\mathcal{M}$ for the Weyl spinor components of generic fermions 
$f_{i}'=(f_{L,i}',f_{R,i}^{\prime \dag})^T$ of generation $i$,
\be
\mathcal{L}\supset (\mathcal{M})_{ij} f_{R,i}' f_{L,j}'+\textrm{H.c.}\,,
\ee
where a sum over repeated indices is implied. 
Following \refsec{sec:models}, the mass matrix takes the generic form
\be\label{eq:fullM}
\mathcal{M}_{Q(L)}= \frac{1}{\sqrt{2}}\left( {\begin{array}{cccc}
y_1 v_h & 0 & 0 & 0 \\
0 & y_2 v_h & 0 & 0 \\
0 & 0 & y_3 v_h & 0 \\
 \lam_{Q(L),1} v_S & \lam_{Q(L),2} v_S & \lam_{Q(L),3} v_S & \sqrt{2}\, m_{Q(L)} 
 \end{array} } \right)\,,
\ee
which can be applied to quarks~($Q=U,D$) and/or to the leptons~($L=E,N$). 
In \refeq{eq:fullM} the SM fermions have been rotated to their diagonal basis already. 

We now diagonalize the $4\times 4$ 
matrix $\mathcal{M}$ ($= \mathcal{M}_{Q},  \mathcal{M}_{L}$) 
by means of 2 unitary matrices 
$\mathcal{U}_L$, $\mathcal{U}_R$,
\be\label{eq:diag}
\mathcal{M}_{\textrm{diag}}=\mathcal{U}_R^{\dag} \mathcal{M}\, \mathcal{U}_L\,,
\ee  
and express the fermion ``gauge'' eigenstates (primed) as function of the physical 
eigenstates (unprimed) as
\bea\label{eq:trans}
f_{L,i}'&=&(\mathcal{U}_L)_{ik}\, f_{L,k}\label{eq:rotL} \\
f_{R,i}'&=&(\mathcal{U}_R)_{ik}^{\ast}\, f_{R,k}\,,\label{eq:rotR} 
\eea
where, again, a sum over repeated indices is implied.

A VL fermion of 
generation $\bar{k}$, charged under the U(1)$_X$ symmetry with $Q_X\neq 0$,
develops a gauge interaction with the $Z'$ boson,
\be\label{eq:gauge}
\mathcal{L}\supset g_X Q_X \left(\bar{f}_{\bar{k}}'\gamma^{\mu} f_{\bar{k}}'\right) \tilde{Z}_{\mu}'\,.
\ee
One can insert Eqs.~(\ref{eq:rotL}), (\ref{eq:rotR}) into \refeq{eq:gauge}
to extract the couplings of 
the $Z'$ to the physical particles. 
One gets
\be\label{eq:mass}
\mathcal{L}\supset \bar{f}_i\gamma^\mu\left(g_L^{ij} P_L+g_R^{ij} P_R\right) f_j\,\tilde{Z}_\mu'\,,
\ee
where
\be
g_L^{ij}=g_X Q_X (\mathcal{U}^{\dag}_L )_{i\bar{k}}\left(\mathcal{U}_L \right)_{\bar{k}j}\,, \qquad 
g_R^{ij}=g_X Q_X (\mathcal{U}^{\dag}_R )_{i\bar{k}}\left(\mathcal{U}_R \right)_{\bar{k}j} \,.
\ee
Note that the specific texture considered in  
\refeq{eq:fullM} leads to $g_R^{ij}=0$ for $i,j\neq \bar{k}$, in agreement with the results of
\refsec{sec:quark} and \refsec{sec:lepts}.

\subsection{Neutral gauge bosons\label{app:vrot}}

In the presence of fermions charged under the U(1)$_Y$ and U(1)$_X$ gauge groups
kinetic mixing of the abelian gauge bosons, $\epsilon$, will be generated in the Lagrangian.

Let us consider
\be\label{eq:kinmix}
\mathcal{L}\supset -\frac{1}{4}W^i_{\mu\nu}W^{i\mu\nu}-\frac{1}{4}B_{\mu\nu}B^{\mu\nu}-\frac{1}{4}X_{\mu\nu}X^{\mu\nu}-\frac{\eps}{2} B_{\mu\nu}X^{\mu\nu},
\ee
where $W^i_{\mu\nu}$, $B_{\mu\nu}$ and $X_{\mu\nu}$ are the field strength tensors of SU(2)$_L$, U(1)$_Y$, and U(1)$_X$, respectively. After both the electroweak and U(1)$_X$ symmetries are spontaneously broken, the three neutral gauge bosons mix at the tree level. Once the kinetic terms are canonically normalized 
the relation between the gauge and physical bases up to $\mathcal{O}(\eps)$ reads\cite{Liu:2017lpo}
\bea
\tilde{Z}_\mu&=&Z_\mu\cos\theta_M + Z'_\mu\sin\theta_M\,,\label{eq:ztilde}\\
\tilde{A}_\mu&=&A_\mu +\eps\cos\theta_W (Z \sin\theta_M-Z'_\mu \cos\theta_M)\,, \label{eq:atilde}\\
\tilde{Z}_\mu'&=&Z_\mu(-\sin\theta_M+\eps\cos\theta_M\sin\theta_W)+Z'_\mu(\cos\theta_M+\eps\sin\theta_M\sin\theta_W)\,,\label{eq:xtilde}
\eea
where we indicate the fields in the unrotated basis with a 
tilde and those in the physical basis without tilde.
Here $\sin\theta_W$ is the Weinberg angle and 
we define the mixing angle $\tan\theta_M=1/(\beta\pm\sqrt{\beta^2+1})$ 
as
\be\label{eq:betaM}
\beta=\frac{m^2_{Z,\textrm{SM}}\left(1-\eps^2\cos^2\theta_W\right)^2-m^2_{Z'}\left(1-\eps^2-\eps^2\sin^2\theta_W\right)}{2\,m^2_{Z'}\,\eps\sin\theta_W\sqrt{1-\eps^2}}\,,\\
\ee
where $m_{Z,\textrm{SM}}^2=(g_2^2+g_Y^2)\, v_h^2/4$ and $m_{Z'}^2= g_X^2 Q_S^2 v_S^2$ like in \refsec{sec:models}.\footnote{Note 
that these expressions have to be further modified at $\mathcal{O}(\epsilon^2)$ 
by the rotation~(\ref{eq:betaM}) -- the full form of the $Z$ and $Z'$ physical masses 
can be found, \textit{e.g.}, in Ref.\cite{Liu:2017lpo}.}
The $+(-)$ sign in the line above \refeq{eq:betaM} applies to the case $m_{Z,\textrm{SM}}>m_{Z'}$ $(m_{Z,\textrm{SM}}<m_{Z'})$.

The flavor-diagonal couplings of the SM gauge 
bosons $\tilde{A}_{\mu}$ and $\tilde{Z}_{\mu}$ to a
fermion $f=(f_L,f_R^{\dag})^T$, 
of electric charge $\tilde{Q}_f$, 
read
\be
\mathcal{L}\supset g_Y\cos\theta_W \tilde{Q}_f \left(\bar{f}\gamma^\mu f\right)\tilde{A}_\mu+\frac{g_2}{\cos\theta_W} \bar{f}\gamma^\mu\left(c_L^{f} P_L+c_R^{f} P_R\right) f \tilde{Z}_\mu \,,
\ee
where
\be
c^f_{L,R}=\left(T_3^{f_{L,R}} - \tilde{Q}_f\sin^2\theta_{W}\right),
\ee
with $T_3^{f_{L,R}}$ denoting the third component of weak isospin. 
From Eqs.~(\ref{eq:ztilde}), (\ref{eq:atilde}) and 
\refeq{eq:betaM} one can read off the coupling of the physical $Z'$ boson to the SM fermions. 
Up to $\mathcal{O}(\eps)$ one obtains
\be\label{eq:SM_coup}
\mathcal{L}\supset
- \eps\, g_Y\cos^2\theta_W \tilde{Q}_f\left(\bar{f}\gamma^\mu f\right)Z'_\mu+
\eps\, g_2\tan\theta_W\frac{m^2_{Z'}}{m^2_{Z'}-m^2_{Z,\textrm{SM}}}\bar{f}\gamma^\mu\left(c_L^fP_L+c_R^f P_R\right) f Z'_\mu \,.
\ee

Due to the flavor construction described in Appendix~\ref{app:frot} the model also features 
interactions of the $\tilde{Z}_{\mu}'$ boson with fermions of different generations $i,j$.
Consider the Lagrangian in \refeq{eq:mass} which, 
to provide a solution for the flavor anomalies, must apply in particular 
to the quarks of the second and third generation, 
and to the leptons of the second generation. After plugging in \refeq{eq:xtilde} one can extract 
the off-diagonal couplings of the physical $Z'$ boson to the fermions. 

\subsection{Scalars\label{app:srot}}
The scalar potential of the models considered in this study takes the form 
\be\label{eq:sca_pot_ap}
V\left(\left|h\right|^2,\left|S\right|^2\right)=-\mu_h^2\, h^{\dag}h+\lam_h\left(h^{\dag}h\right)^2+\mu_S^2\, S^{\dag}S
+\lam_S \left(S^{\dag}S\right)^2+\lam_{hS}\left(S^{\dag}S\right)\left(h^{\dag}h\right)\,,
\ee
where 
\be
h:(\mathbf{1},\mathbf{2},1/2,0)\,,
\ee
is the SM Higgs doublet. 

After the breaking of the electroweak and U(1)$_X$ symmetries, respectively by the vev's $v_h$ and $v_S$,
the mass matrix for the real part of the neutral 
component of the Higgs doublet $h$ and the real part of the scalar $S$ takes the form
\be\label{eq:fullMSc}
m^2_H= \left( 
{\begin{array}{cc}
-\mu_h^2+ 3 \lam_h v_h^2 + \frac{1}{2} \lam_{hS} v_S^2 & \lam_{hS} v_S v_h  \\
\lam_{hS} v_S v_h & \mu_S^2 + 3\lam_S v_S^2 + \frac{1}{2} \lam_{hS} v_h^2   \\
\end{array} } \right)\,.
\ee

Equation~(\ref{eq:fullMSc}) can be diagonalized by an orthogonal matrix $\mathcal{Z}_H$ such that
\be
(m^2_H)_{\textrm{diag}}=\mathcal{Z}_H m^2_H \mathcal{Z}_H^T\,.
\ee
The mixing angle $\alpha_H$ parameterizing the matrix $\mathcal{Z}_H$ is given by 
\be\label{eq:mixan}
\sin\alpha_H=\frac{\lam_{hS} \,v_h \,v_S}{2 \sqrt{\lam_h^2 v_h^4 + \lam_{hS}^2 v_h^2 v_S^2 - 2 \lam_h \lam_S v_h^2 v_S^2 + \lam_S^2 v_S^4}}\,,
\ee
and the physical masses of the scalars 
$H_1$ and $H_2$ can be expressed in terms of the parameters in \refeq{eq:sca_pot_ap} as
\bea
m_{H_1}&=&\sqrt{\lam_h v_h^2 + \lam_S v_S^2 -\sqrt{\lam_h^2 v_h^4 + \lam_{hS}^2 v_h^2 v_S^2 - 2 \lam_h \lam_S v_h^2 v_S^2 + \lam_S^2 v_S^4}}\,,\nonumber\\
m_{H_2}&=&\sqrt{\lam_h v_h^2 + \lam_S v_S^2 +\sqrt{\lam_h^2 v_h^4 + \lam_{hS}^2 v_h^2 v_S^2 - 2 \lam_h \lam_S v_h^2 v_S^2 + \lam_S^2 v_S^4}}\,.\label{eq:maS}
 \eea

\section{RGEs in the presence of two U(1) group factors \label{app:RGEsKM}}
Let us consider a gauge theory with a symmetry group U(1)$_Y\times$U(1)$_X$, with the corresponding gauge couplings denoted as 
$g_Y$ and $g_X$. If the theory contains a fermion $f$ transforming under both symmetry factors with charges $Q_Y$, $Q_X$,
kinetic mixing $\epsilon$ is generated between the two abelian groups. The gauge part of the Lagrangian takes the form
\bea\label{eq:lagE}
\mathcal{L}&\supset& -\frac{1}{4}B_{\mu\nu}B^{\mu\nu}-\frac{1}{4}X_{\mu\nu}X^{\mu\nu}-\frac{\eps}{2} B_{\mu\nu}X^{\mu\nu}\nonumber\\
& &\qquad  +i\bar{f}\left(\partial^\mu-i g_Y Q_Y \tilde{B}^\mu-i g_X Q_X \tilde{X}^\mu\right)\gamma_\mu f\,,
\eea
where we indicate with 
$\tilde{B}^\mu$ and $\tilde{X}^\mu$ the gauge bosons of U(1)$_Y$ and U(1)$_X$, respectively, 
and $B_{\mu\nu}$ and $X_{\mu\nu}$ are the corresponding field strength tensors. 

In order to modify the RGEs with trans-Planckian corrections linear in the gauge couplings it 
is convenient to work in a basis in which the gauge fields are canonically normalized. This can be achieved by a rotation\cite{Holdom:1985ag,Babu:1996vt}
\be
\begin{pmatrix}
\tilde{B}^\mu\\
\tilde{X}^\mu
\end{pmatrix}
=\begin{pmatrix}
1 && -\eps/\sqrt{(1-\eps^2)} \\
0 && 1/\sqrt{(1-\eps^2)}  
\end{pmatrix}
\begin{pmatrix}
V^\mu\\
D^\mu
\end{pmatrix}\,,
\ee
which parameterizes the gauge interaction vertices of Lagrangian~(\ref{eq:lagE}) in terms of a ``visible'' gauge boson $V^{\mu}$
and a ``dark'' gauge boson $D^{\mu}$:
\be\label{mix:vertex}
(Q_Y\,Q_X)
\left(\begin{array}{cc}
g_{Y} & 0 \\ 0  & g_{X} 
\end{array} \right)
\left(\begin{array}{c}
\tilde{B}^\mu\\ \tilde{X}^\mu
\end{array} \right)\quad \to \quad
(Q_Y\,Q_X)
\left(\begin{array}{cc}
g_{V} & g_\eps \\ 0  & g_{D} 
\end{array} \right)
\left(\begin{array}{c}
V^\mu\\ D^\mu
\end{array} \right)\,.
\ee
The elements $g_V$, $g_D$, and $g_\eps$ 
are related to the original gauge couplings $g_Y$, $g_X$ and the kinetic mixing $\eps$ as
\be\label{eq:gcoups}
g_V=g_Y,\quad g_D=\frac{g_X}{\sqrt{1-\eps^2}},\quad g_\eps=-\frac{\eps g_Y}{\sqrt{1-\eps^2}}\,.
\ee


Generic sub-Planckian RGEs in the $g_V$, $g_D$ and $g_\eps$ basis are given by\cite{Babu:1996vt}
\bea
\frac{dg_V}{dt}&=&\frac{1}{16\pi^2}\frac{2}{3} d(R_{F3}) d(R_{F2})\, Q_Y^2\,g_V^3\label{eq:by}\\
\frac{dg_D}{dt}&=&\frac{1}{16\pi^2}\frac{2}{3} d(R_{F3}) d(R_{F2}) \left(Q_Y^2g_\eps^2+2Q_Y Q_Xg_\eps g_D+Q_X^2 g_D^2\right)g_D \label{eq:bd} \\
\frac{dg_\eps}{dt}&=&\frac{1}{16\pi^2}\frac{2}{3} d(R_{F3}) d(R_{F2}) \left[Q_Y^2 (2 g_V^2\, g_\eps + g_\eps^3)+2Q_Y Q_Xg_D (g_D^2 + g_\eps^2)+Q_X^2 g_D^2g_\eps\right],\label{eq:be}
\eea
where $d(R_{F3})$ is the dimension of the SU(3)$_\textrm{c}$ representation $R_{F3}$ under which the fermion $f$ transforms and $d(R_{F2})$ 
is the corresponding dimension of the SU(2)$_L$ representation $R_{F2}$.

While Eqs.~(\ref{eq:by}) and (\ref{eq:bd}) are multiplicative in the respective gauge couplings, 
\refeq{eq:be} shows that $g_{\epsilon}$ may be generated additively, 
if and only if both charges $Q_Y$, $Q_X$
are different from zero. It is easy to invert \refeq{eq:gcoups} and read off the kinetic mixing from the running couplings:
\be\label{eq:epsilon}
\epsilon=\frac{g_{\epsilon}}{\sqrt{g_{\epsilon}^2+g_V^2}}\,.
\ee

In the presence of multiple fields charged under U(1)$_Y$, U(1)$_X$, or both, 
\refeqs{eq:by}{eq:be} can be rewritten in compact form,
\bea
\frac{dg_V}{dt}&=&\frac{1}{16\pi^2} \widetilde{Q}_Y\,g_V^3\\
\frac{dg_D}{dt}&=&\frac{1}{16\pi^2} \left( \widetilde{Q}_Y g_\eps^2+2\, \widetilde{Q}_{YX}g_\eps g_D+ \widetilde{Q}_Xg_D^2\right)g_D\\
\frac{dg_\eps}{dt}&=&\frac{1}{16\pi^2} \left[ \widetilde{Q}_Y \left(2\, g_V^2 g_\eps + g_\eps^3\right)+2\, \widetilde{Q}_{YX} g_D \left(g_V^2 + g_\eps^2\right)+ \widetilde{Q}_X g_D^2 g_\eps\right]\,,\label{eq:be_comp}
\eea
where the one-loop coefficients $\widetilde{Q}_Y$, $\widetilde{Q}_X$ and $\widetilde{Q}_{YX}$ are defined as
\bea
\widetilde{Q}_Y&=&\frac{2}{3}\sum_i d(R_{i3}) d(R_{i2}) Q_{Yi}^2+\frac{1}{3}\sum_j d(R_{j3}) d(R_{j2}) Q_{Yj}^2 \label{eq:qu1} \\
\widetilde{Q}_X&=&\frac{2}{3}\sum_i d(R_{i3}) d(R_{i2}) Q_{Xi}^2+\frac{1}{3}\sum_j d(R_{j3}) d(R_{j2}) Q_{Xj}^2 \label{eq:qux} \\
\widetilde{Q}_{YX}&=&\frac{2}{3}\sum_i d(R_{i3}) d(R_{i2}) Q_{Yi}Q_{Xi}+\frac{1}{3}\sum_j d(R_{j3}) d(R_{j2}) Q_{Yj}Q_{Xj}\,,\label{eq:qu1x}
\eea
where $Q_{Yi(j)}$ and $Q_{Xi(j)}$ are the abelian charges of a fermion (scalar) particle of index $i$ ($j$) under
U(1)$_Y$ and U(1)$_X$, respectively, 
and the sums run over all the fermions and scalars in the theory. 

\section{RGEs of the gauge-Yukawa-quartic system\label{app:RGEs}}

The elements of the SM and NP Yukawa matrices are all subject to modifications 
due to RGE running. When working in the mass basis the scale dependence of the rotation matrices 
is encapsulated in the running of the elements of the CKM matrix, see, \textit{e.g.}, Appendices~A and B of Ref.\cite{Kowalska:2020gie} and references therein for a discussion. Following the procedure 
detailed there, we compute the RGEs in the flavor basis with \texttt{SARAH v4.14.0}\cite{Staub:2010jh} and \texttt{RGBeta}\cite{Thomsen:2021ncy} and then transform them to the quark mass basis. We work in the ``down-origin'' approach, in which the NP Yukawa couplings of the up-type quarks are related to the more 
``fundamental'' down-type Yukawa couplings $\lam_{Q,i}$ by a CKM rotation. We then employ the 2-family approximation, 
in which the CKM matrix is orthogonal and defined by one independent parameter, 
\be
V_{22}=V_{33}\,,\quad \quad V_{23}=-V_{32}=\sqrt{1-V_{33}{}^2}\,.
\ee

In the following we present the trans-Planckian RGEs of the models defined in \reftable{tab:mod_kin} for $Q_S=-1$. 
All parameters that are not shown explicitly are
considered to be zero and relevant at the UV fixed point. We restrict our analysis to one loop 
and consider only trans-Planckian corrections to the 
RGEs that are linear in the coupling constants, neglecting the effects of higher order terms.  
Note that $t=\log k$.

\subsection{Model 1 (VL leptons)\label{app:RGEs_model1}}

\subsubsection{Gauge sector Model~1A}
\medskip
\be
\frac{dg_3}{dt}=-\frac{17}{3}\frac{g_3^3}{16\pi^2}-f_g g_3
\ee
\be
\frac{dg_2}{dt}=-\frac{1}{2}\frac{g_2^3}{16\pi^2}-f_g g_2
\ee
\be
\frac{dg_Y}{dt}=\frac{139}{18}\frac{g_Y^3}{16\pi^2}-f_g g_Y
\ee
\be
\frac{dg_D}{dt}=\frac{1}{16\pi^2}\left(11 g^2_D+\frac{139}{18}g^2_\eps\right)g_D-f_g g_D
\ee
\be
\frac{dg_\eps}{dt}=\frac{1}{16\pi^2}\left(11 g_D^2 g_\eps + \frac{139}{9} g_Y^2 g_\eps + \frac{139}{18} g_\eps^3\right)-f_g g_\eps
\ee

\subsubsection{Gauge sector Model~1B}
\medskip
\be
\frac{dg_3}{dt}=-\frac{17}{3}\frac{g_3^3}{16\pi^2}-f_g g_3
\ee
\be
\frac{dg_2}{dt}=-\frac{1}{2}\frac{g_2^3}{16\pi^2}-f_g g_2
\ee
\be
\frac{dg_Y}{dt}=\frac{139}{18}\frac{g_Y^3}{16\pi^2}-f_g g_Y
\ee
\be
\frac{dg_D}{dt}=\frac{1}{16\pi^2}\left(11 g^2_D+\frac{139}{18}g^2_\eps-\frac{16}{3}g_Dg_\eps\right)g_D-f_g g_D
\ee
\be
\frac{dg_\eps}{dt}=\frac{1}{16\pi^2}\left(11 g_D^2 g_\eps + \frac{139}{9} g_Y^2 g_\eps + \frac{139}{18} g_\eps^3-\frac{16}{3}g_Dg^2_Y-\frac{16}{3}g_Dg^2_\eps\right)-f_g g_\eps
\ee

\medskip
\subsubsection{Yukawa and scalar sector}
\medskip
\begin{multline}
\frac{d y_t}{dt}=\frac{1}{16\pi^2}\left[3 y_b^2+\frac{9}{2} y_t^2-\frac{17}{12}g_Y^2-\frac{17}{12}g_\epsilon^2-\frac{9}{4}g_2^2-8 g_3^2-\frac{3}{2}V_{33}{}^2 y_b^2\right.\\
\left.+\frac{1}{2}V_{32}{}^2(\lam_{Q,2})^2+ V_{32}V_{33}\lam_{Q,3}\lam_{Q,2}
+\frac{1}{2}V_{33}{}^2(\lam_{Q,3})^2 \right]y_t-f_y y_t
\end{multline}
\begin{equation}
\frac{d y_b}{dt}=\frac{1}{16\pi^2}\left[\frac{9}{2} y_b^2+3 y_t^2-\frac{5}{12}g_Y^2-\frac{5}{12}g_\epsilon^2-\frac{9}{4}g_2^2-8 g_3^2 -\frac{3}{2}V_{33}{}^2 y_t^2
+\frac{1}{2}(\lam_{Q,3})^2 \right]y_b-f_y y_b
\end{equation}
\begin{multline}
    \frac{d\lam_{Q,2}}{dt}=\frac{1}{16\pi^2}\left\{\left[7(\lam_{Q,2})^2+ \frac{13}{2}(\lam_{Q,3})^2+2(\lam_{L,2})^2+\frac{1}{2}y_t^2 V_{32}{}^2-\frac{9}{2}g_2^2-8 g_3^2\right.\right.\\
\left.\left.-\frac{1}{6}g_Y^2-\frac{1}{6}g_\epsilon^2-3 g_D^2 +  g_D g_\epsilon  \right]\lam_{Q,2}+2 y_t^2 V_{32}V_{33}\lam_{Q,3}\right\}-f_y \lam_{Q,2}
\end{multline}
\begin{multline}
    \frac{d\lam_{Q,3}}{dt}=\frac{1}{16\pi^2}\left\{\left[\frac{15}{2}(\lam_{Q,2})^2+ 7(\lam_{Q,3})^2+2(\lam_{L,2})^2+\frac{1}{2}y^2_b+\frac{1}{2}y_t^2 V_{33}{}^2 -\frac{9}{2}g_2^2-8 g_3^2\right.\right.\\
\left.\left.-\frac{1}{6}g_Y^2-\frac{1}{6}g_\epsilon^2-3 g_D^2 + g_D g_\epsilon \right]\lam_{Q,3}- y_t^2 V_{32}V_{33}\lam_{Q,2}\right\}-f_y \lam_{Q,3}
\end{multline}
\begin{multline}
    \frac{d\lam_{L,2}}{dt}=\frac{1}{16\pi^2}\left[6(\lam_{Q,2})^2+ 6(\lam_{Q,3})^2+3(\lam_{L,2})^2-\frac{9}{2}g_2^2-\frac{3}{2}g_Y^2-\frac{3}{2}g_\epsilon^2-3 g_D^2\right.\\
   +  3 g_D g_\epsilon \Big]\lam_{L,2}-f_y \lam_{L,2}
\end{multline}
\begin{multline}
    \frac{d|V_{33}|}{dt}=\frac{V_{23}}{16\pi^2}
    \left[-\frac{3}{2}V_{23}V_{33}y_b^2
    +\frac{1}{2}\left(V_{22}V_{32}(\lam_{Q,2})^2\right.\right.\\
    \left.\left.+V_{22}V_{33}\lam_{Q,2}\lam_{Q,3}+V_{23}V_{32}\lam_{Q,2}\lam_{Q,3}+V_{23}V_{33}(\lam_{Q,3})^2 \right) \right]\\
-\frac{V_{32}}{16\pi^2}\left[\frac{3}{2}V_{32}V_{33}y_t^2
    -\frac{1}{2}\lam_{Q,2}\lam_{Q,3}\right]
\end{multline}
\begin{multline}
    \frac{d \lambda_h}{dt}=\frac{1}{16\pi^2}\left[\frac{3}{8}(g_Y^2+g_\eps^2)^2+\frac{3}{4}(g_Y^2+g_\eps^2)g_2^2+\frac{9}{8}g_2^4-3g_Y^2\lam_h-3g_\eps^2\lam_h-9g_2^2\lam_h\right.\\ 
    \left.+ 24 \lam_h^2 +\lam_{hS}^2 + 12y_b^2\lam_h + 12y_t^2\lam_h - 6 y_b^4 - 6 y_t^4 \right] - f_\lam \lam_h
\end{multline}
\begin{multline}
    \frac{d \lambda_S}{dt}=\frac{1}{16\pi^2}\left[6g_D^4g_Y^2/(g_Y^2+g^2_\eps)+ 2\lam_{hS}^2-12g_D^2\lam_S + 20\lam_S^2+8(\lam_{L,2})^2\lam_S-4(\lam_{L,2})^4 \right.\\
    \left. +24(\lam_{Q,2})^2\lam_S +24 (\lam_{Q,3})^2\lam_S -12\left( (\lam_{Q,2})^2 +(\lam_{Q,3})^2\right)^2\right] - f_\lam \lam_S
\end{multline}
\begin{multline}
    \frac{d \lambda_{hS}}{dt}=\frac{1}{16\pi^2}\left[-\frac{3}{2}g_Y^2\lam_{hS} -\frac{9}{2} g_2^2\lam_{hS}-6 g_D^2\lam_{hS}+12\lam_{h}\lam_{hS}+4 \lam_{hS}^2 + 8\lam_{hS}\lam_{S}\right.\\
    \left. +4 (\lam_{L,2})^2 \lam_{hS} + 12(\lam_{Q,2})^2 \lam_{hS} + 12(\lam_{Q,3})^2 \lam_{hS} + 6y_b^2 \lam_{hS} + 6 y_t^2 \lam_{hS}\right.\\
    \left.- 12 y_b^2 (\lam_{Q,3})^2 - 12 y_t^2V_{32}{}^2(\lam_{Q,2})^2-12y^2_t V_{33}{}^2(\lam_{Q,3})^2\right.\\
    \left.-12y^2_tV_{32}V_{33}\lam_{Q,2}\lam_{Q,3}\right] - f_\lam \lam_{hS}\,.
\end{multline}

\subsection{Model~2 ($L_\mu-L_{\tau}$)\label{app:RGEs_mutau}}
\medskip
\be
\frac{dg_3}{dt}=-\frac{17}{3}\frac{g_3^3}{16\pi^2}-f_g g_3
\ee
\be
\frac{dg_2}{dt}=-\frac{7}{6}\frac{g_2^3}{16\pi^2}-f_g g_2
\ee
\be
\frac{dg_Y}{dt}=\frac{127}{18}\frac{g_Y^3}{16\pi^2}-f_g g_Y
\ee
\be
\frac{dg_D}{dt}=\frac{1}{16\pi^2}\left[\frac{37}{3}g^2_D-\frac{8}{3}g_Dg_\epsilon+\frac{127}{18}g^2_\epsilon\right]g_D-f_g g_D
\ee
\be
\frac{dg_\epsilon}{dt}=\frac{1}{16\pi^2}\left[\frac{37}{3}g^2_Dg_\epsilon-\frac{8}{3}g_Dg^2_\epsilon-\frac{8}{3}g_Dg^2_Y+\frac{127}{18}g^3_\epsilon+\frac{127}{9}g_\epsilon g^2_Y\right]-f_g g_\epsilon
\ee
\begin{multline}
\frac{d y_t}{dt}=\frac{1}{16\pi^2}\left[3 y_b^2+\frac{9}{2} y_t^2-\frac{17}{12}g_Y^2-\frac{17}{12}g_\epsilon^2-\frac{9}{4}g_2^2-8 g_3^2-\frac{3}{2}V_{33}{}^2 y_b^2\right.\\
\left.+\frac{1}{2}\left(V_{32}{}^2(\lam_{Q,2})^2+2 V_{32}V_{33}\lam_{Q,3}\lam_{Q,2}
+V_{33}{}^2(\lam_{Q,3})^2\right) \right]y_t-f_y y_t
\end{multline}
\begin{equation}
\frac{d y_b}{dt}=\frac{1}{16\pi^2}\left[\frac{9}{2} y_b^2+3 y_t^2-\frac{5}{12}g_Y^2-\frac{5}{12}g_\epsilon^2-\frac{9}{4}g_2^2-8 g_3^2 -\frac{3}{2}V_{33}{}^2 y_t^2
+\frac{1}{2}(\lam_{Q,3})^2 \right]y_b-f_y y_b
\end{equation}
\begin{multline}
    \frac{d\lam_{Q,2}}{dt}=\frac{1}{16\pi^2}\left\{\left[7(\lam_{Q,2})^2+ \frac{13}{2}(\lam_{Q,3})^2-\frac{1}{6}g_Y^2-\frac{1}{6}g_\epsilon^2-3 g_D^2 + g_D g_\epsilon-\frac{9}{2}g_2^2-8 g_3^2\right.\right.\\
\left.\left.+\frac{1}{2}y_t^2 V_{32}{}^2  \right]\lam_{Q,2}+2 y_t^2 V_{32}V_{33}\lam_{Q,3}\right\}-f_y \lam_{Q,2}
\end{multline}
\begin{multline}
    \frac{d\lam_{Q,3}}{dt}=\frac{1}{16\pi^2}\left\{\left[\frac{15}{2}(\lam_{Q,2})^2+ 7(\lam_{Q,3})^2-\frac{1}{6}g_Y^2-\frac{1}{6}g_\epsilon^2-3 g_D^2 + g_D g_\epsilon-\frac{9}{2}g_2^2-8 g_3^2\right.\right.\\
\left.\left.+\frac{1}{2}y^2_b+\frac{1}{2}y_t^2 V_{33}{}^2  \right]\lam_{Q,3}- y_t^2 V_{32}V_{33}\lam_{Q,2}\right\}-f_y \lam_{Q,3}
\end{multline}
\begin{multline}
    \frac{d|V_{33}|}{dt}=\frac{V_{23}}{16\pi^2}
    \left[-\frac{3}{2}V_{23}V_{33}y_b^2
    +\frac{1}{2}\left(V_{22}V_{32}(\lam_{Q,2})^2\right.\right.\\
    \left.\left.+V_{22}V_{33}\lam_{Q,2}\lam_{Q,3}+V_{23}V_{32}\lam_{Q,2}\lam_{Q,3}+V_{23}V_{33}(\lam_{Q,3})^2 \right) \right]\\
-\frac{V_{32}}{16\pi^2}\left[\frac{3}{2}V_{32}V_{33}y_t^2
    -\frac{1}{2}\lam_{Q,2}\lam_{Q,3}\right]
\end{multline}
\begin{multline}
    \frac{d \lambda_h}{dt}=\frac{1}{16\pi^2}\left[\frac{3}{8}(g_Y^2+g_\eps^2)^2+\frac{3}{4}(g_Y^2+g_\eps^2)g_2^2+\frac{9}{8}g_2^4-3g_Y^2\lam_h-3g_\eps^2\lam_h-9g_2^2\lam_h\right.\\ 
    \left.+ 24 \lam_h^2 +\lam_{hS}^2 + 12y_b^2\lam_h + 12y_t^2\lam_h - 6 y_b^4 - 6 y_t^4 \right] - f_\lam \lam_h
\end{multline}
\begin{multline}
    \frac{d \lambda_S}{dt}=\frac{1}{16\pi^2}\left[6g_D^4g_Y^2/(g_Y^2+g^2_\eps)+ 2\lam_{hS}^2-12g_D^2\lam_S + 20\lam_S^2+24(\lam_{Q,2})^2\lam_S +24 (\lam_{Q,3})^2\lam_S  \right.\\
    \left.-12\left( (\lam_{Q,2})^2 +(\lam_{Q,3})^2\right)^2\right] - f_\lam \lam_S
\end{multline}
\begin{multline}
    \frac{d \lambda_{hS}}{dt}=\frac{1}{16\pi^2}\left[-\frac{3}{2}g_Y^2\lam_{hS} -\frac{9}{2} g_2^2\lam_{hS}-6 g_D^2\lam_{hS}+12\lam_{h}\lam_{hS}+4 \lam_{hS}^2 + 8\lam_{hS}\lam_{S}\right.\\
    \left. + 12(\lam_{Q,2})^2 \lam_{hS} + 12(\lam_{Q,3})^2 \lam_{hS} + 6y_b^2 \lam_{hS} + 6 y_t^2 \lam_{hS}\right.\\
    \left.- 12 y_b^2 (\lam_{Q,3})^2 - 12 y_t^2V_{32}{}^2(\lam_{Q,2})^2-12y^2_t V_{33}{}^2(\lam_{Q,3})^2\right.\\
    \left.-12y^2_tV_{32}V_{33}\lam_{Q,2}\lam_{Q,3}\right] - f_\lam \lam_{hS}\,.
\end{multline}

\section{Neutrino masses and dark matter from feebly interacting particles\label{app:feeble_ext}}

As anticipated in \refsec{sec:stabil}, the impact of additional NP on the RGEs of Appendix~\ref{app:RGEs}
can be minimized or canceled altogether if one assumes that the SM extension addressing phenomena beyond the flavor anomalies mostly comprises feebly interacting particles. 
As a simple example, let us introduce a model of sterile-neutrino dark matter, adopted from Refs.\cite{Kusenko:2006rh,Petraki:2007gq,Frigerio:2014ifa}. We add to the particle content of \refsec{sec:models} four gauge-singlet fields,
\be
S_1,\, \nu_{i=1,2,3}:(\textbf{1},\textbf{1},0,0)\,, 
\ee
where $S_1$ is a real scalar and $\nu_{i=1,2,3}$ are Weyl spinors (``right-handed'' neutrinos).  The Lagrangian admits new Yukawa couplings and a Majorana mass term, which may (or may not) be a consequence of $S_1$ developing a vev. One writes 
\be\label{eq:maj}
\mathcal{L}\supset -y_{\nu,ij}\, \nu_{i} (h^c)^{\dag} l_j - Y_{S,ij} \, S_1 \nu_{i} \nu_{j}  -\frac{1}{2} M_{N,ij}\, \nu_{i} \nu_{j}
 + \textrm{H.c.}\,,
\ee
where a sum over flavor indices is implied. The scalar potential will also be extended with dimensionful and dimensionless couplings of the field $S_1$, with the latter remaining small as explained in \refsec{sec:scal_FP}. 

It was shown in Refs.\cite{Kusenko:2006rh,Petraki:2007gq,Frigerio:2014ifa} that in this scenario
a reliable process for the production of sterile-neutrino dark matter is provided by 
the freeze-in mechanism\cite{McDonald:2001vt,Choi:2005vq,Kusenko:2006rh,Petraki:2007gq,Hall:2009bx}. One assumes that the couplings 
between the visible sector and the dark matter are very small,
so that the latter never reaches thermal equilibrium. The observed relic abundance is eventually generated by the decay
of the $S_1$ particle, which remains in thermal equilibrium, into dark-matter final states. One gets, approximately, 
\be\label{eq:frin}
\Omega h^2 \approx 0.12 \left(\frac{Y_S}{10^{-8}} \right)^2 \left( \frac{M_N}{10^{-8}\, m_{S_1}} \right)\,,
\ee
where we have dropped for simplicity the generation indices.

From \refeq{eq:frin}, one expects a feeble Yukawa coupling, $Y_S\ll 1$, and 
a dark matter mass  by several orders of magnitude smaller than the mass of the decaying particle $S_1$. Assuming, for example, that $m_{S_1}$ lies at collider scales, 
dark matter may be produced via freeze-in if the active neutrino mass were generated in a \textit{light} ``see-saw'' mechanism, with a feeble Yukawa interaction, $y_{\nu}\ll 1$, and a correspondingly small Majorana mass, $M_N\approx 1 \kev-1\mev$.

As some of us have demonstrated in Ref.\cite{Kowalska:2022ypk}, this simple scenario 
for dark matter and neutrino masses can originate
dynamically from a UV completion with AS. The embedding does not affect the TeV-scale phenomenology, 
which can be trusted to a very good approximation. In the next few paragraphs we show that this is the case for the models investigated in this work.

Let us limit our analysis for simplicity to Model~1A,\textit{a}. In the trans-Planckian regime, the sizeable part of the Yukawa coupling RGEs
can be extracted from Appendix~\ref{app:RGEs_model1}, with some modifications due to the terms in \refeq{eq:maj}. Neglecting the neutrino generation indices one writes
\bea
\frac{d y_t}{dt}&\simeq& \frac{1}{16 \pi^2}\left[\frac{9}{2} y_t^2 -\frac{17}{12} g_Y^2+\frac{1}{2}\lam_{Q,i}^2+y_{\nu}^2  \right]y_t - f_y\, y_t \label{eq:newyt} \\ 
\frac{d y_{\nu}}{dt}&\simeq& \frac{1}{16 \pi^2}\left[3 y_t^2 -\frac{9}{12} g_Y^2+\frac{1}{2}\lam_{L,2}^2+2 Y_S^2 +\frac{5}{2} y_{\nu}^2  \right]y_{\nu}-f_y\, y_{\nu}\label{eq:ynu}\\
\frac{d \lam_{L,2}}{dt}&\simeq& \frac{1}{16 \pi^2}\left[6\lam_{Q,i}^2+3 \lam_{L,2}^2-\frac{3}{2} g_Y^2 -3 g_D^2+\frac{1}{2} y_{\nu}^2 \right]\lam_{L,2}-f_y\, \lam_{L,2}\\
\frac{d Y_S}{dt}&\simeq& \frac{1}{16 \pi^2}\left[6 Y_S^2 +2 y_{\nu}^2 \right]Y_S-f_y\, Y_S\,,\label{eq:ys}
\eea
while all other RGEs remain unaltered with respect to Appendix~\ref{app:RGEs_model1}. Following Ref.\cite{Kowalska:2022ypk}, one can show that if the RGE system admits an IR-attractive
fixed point with $y_{\nu}^{\ast}= 0$ an arbitrarily small neutrino Yukawa coupling can be generated dynamically along the trans-Planckian flow, down from additional UV attractive fixed points. It is easy to infer from \refeq{eq:ynu} that such 
an IR-attractive Gaussian solution exists in Model~1A,\textit{a}, if $f_y=0.0025$ like in \reftable{tab:FPall}. 

Even more importantly, the stability matrix shows that the system (\ref{eq:newyt})-(\ref{eq:ys})
admits the exact same IR-attractive fixed points as those shown in \reffig{fig:flow}(a), and that $Y_S^{\ast}=0$ follows a relevant direction for $f_y=0.0025$, so that the fixed point can be connected to any desired value in the IR.
We conclude that a small neutrino mass and the correct relic abundance of dark matter can emerge from Eqs.~(\ref{eq:maj}), (\ref{eq:frin}) without any impact on the fixed points predictions of \refsec{sec:gayusys}.

\bibliographystyle{JHEP}
\bibliography{mybib}

\end{document}

%% file: macros.tex
\newcommand{\newc}{\newcommand*}

\long\def\begincomment#1\endcomment{%
        \begingroup\sf\baselineskip12pt#1\endgroup}

\newc{\etal}{\textrm{et al.}} 
\newc{\eg}{\textrm{e.g.}} 
\newc{\ie}{\textrm{i.e.}}
\newc{\etc}{\textrm{etc}}
\newc\vs{\textrm{vs.}}
\newc{\cl}{\rm {C.L.}}

\newc{\ev}{\ensuremath{\,\mathrm{eV}}}
\newc{\kev}{\ensuremath{\,\mathrm{keV}}}
\newc{\mev}{\ensuremath{\,\mathrm{MeV}}}
\newc{\gev}{\ensuremath{\,\mathrm{GeV}}}
\newc{\tev}{\ensuremath{\,\mathrm{TeV}}}
\newc{\MeV}{\mev} 
\newc{\TeV}{\tev}
\newc{\invpb}{\ensuremath{/\text{pb}}}
\newc{\invfb}{\ensuremath{/\text{fb}}}
\newc\nb{\ensuremath{\,\mathrm{nb}}} \newc\pb{\ensuremath{\,\mathrm{pb}}} \newc\fb{\ensuremath{\,\mathrm{fb}}}
\newc\pc{\ensuremath{\,\mathrm{pc}}}
\newc\kpc{\ensuremath{\,\mathrm{kpc}}}
\newc\mpc{\ensuremath{\,\mathrm{Mpc}}}
\newc\ps{\ensuremath{\,\mathrm{ps}}} 
\newc\cmeter{\ensuremath{\,\mathrm{cm}}} 
\newc\meter{\ensuremath{\,\mathrm{m}}} 
\newc\kmeter{\ensuremath{\,\mathrm{km}}}
\newc\second{\ensuremath{\,\mathrm{s}}}
\newc\msecond{\ensuremath{\,\mathrm{ms}}}
\newc\nsecond{\ensuremath{\,\mathrm{ns}}}
\newc\psecond{\ensuremath{\,\mathrm{ps}}}

\newc{\chisqmin}{\ensuremath{\chi^2_{\mathrm{min}}}}
\newc{\Delchisq}{\ensuremath{\Delta\chi^2}}
\newc{\chisq}{\ensuremath{\chi^2}}
\newc{\like}{\ensuremath{\mathcal{L}}}

\newc\lsim{\ensuremath{\mathrel{\rlap{\lower4pt\hbox{\hskip1pt$\sim$}}\raise1pt\hbox{$<$}}}}
\newc\gsim{\ensuremath{\mathrel{\rlap{\lower4pt\hbox{\hskip1pt$\sim$}}\raise1pt\hbox{$>$}}}}
\newc{\VEV}[1]{\ensuremath{\langle #1 \rangle}}
\newc{\dl}{\ensuremath{\stackrel{\leftarrow}{D}}}
\newc{\dr}{\ensuremath{\stackrel{\rightarrow}{D}}}

\newc{\bcenter}{\begin{center}}   
\newc{\ecenter}{\end{center}}
\newc{\bfl}{\begin{flushleft}}    
\newc{\efl}{\end{flushleft}}
\newc{\bfr}{\begin{flushright}}   
\newc{\efr}{\end{flushright}}

\newc{\bi}{\begin{itemize}}
\newc{\ei}{\end{itemize}}
\newc{\bed}{\begin{description}}
\newc{\eed}{\end{description}}
\newc{\ben}{\begin{enumerate}}
\newc{\een}{\end{enumerate}}

\newc{\be}{\begin{equation}}
\newc{\ee}{\end{equation}}
\newc{\bea}{\begin{eqnarray}}
\newc{\eea}{\end{eqnarray}}
\newc{\bfle}{\begin{flalign}}
\newc{\efle}{\end{flalign}}
\newc{\ra}{\rightarrow}

\newc{\alphas}{\ensuremath{\alpha_s}}
\newc{\alphatwo}{\ensuremath{\alpha_2}}
\newc{\alphaone}{\ensuremath{\alpha_1}}
\newc{\alphai}[1]{\ensuremath{\alpha_{#1}}}
\newc{\alphaem}{\ensuremath{\alpha_{\mathrm{em}}}}
\newc{\alphaeff}{\ensuremath{\alpha_{\mathrm{eff}}}}
\newc{\sineff}{\ensuremath{\sin \theta_{\mathrm{eff}}}}
\newc{\sinsqeff}{\ensuremath{\sin^2 \theta_{\mathrm{eff}}}}
\newc{\dalphahad}{\ensuremath{\Delta \alpha_{\mathrm{had}}}}
\newc{\yt}{\ensuremath{h_t}} \newc{\yb}{\ensuremath{h_b}} \newc{\ytau}{\ensuremath{h_{\tau}}}
\newc\mz{\ensuremath{M_Z}} 
\newc\mw{\ensuremath{m_W}}
\newc\mZ{\mz}        \newc\mW{\mw}
\newc\mhsm{\ensuremath{ m_{H_{\mathrm{SM}}}}}
\newc{\mtop}{\ensuremath{ m_t}}               \newc{\mtpole}{\ensuremath{ M_t}}
\newc{\mbottom}{\ensuremath{ m_b}} 
\newc{\mtau}{\ensuremath{ m_{\tau}}}
\newc{\mt}{\mtpole}
\newc{\mb}{\mbottom} 
\newc{\rtwogg}{\ensuremath{R_{h_2}(\gamma\gamma)}}
\newc{\rtwozz}{\ensuremath{R_{h_2}(ZZ)}}
\newc{\ronegg}{\ensuremath{R_{h_1}(\gamma\gamma)}}
\newc{\ronezz}{\ensuremath{R_{h_1}(ZZ)}}
\newc{\rsiggg}{\ensuremath{R_{h_\textrm{sig}}(\gamma\gamma)}}
\newc{\rsigzz}{\ensuremath{R_{h_\textrm{sig}}(ZZ)}}
\newc{\llbar}{\ensuremath{\ell\bar{\ell}}}
\newc{\tauptaum}{\ensuremath{ \tau^+\tau^-}}
\newc{\qqbar}{\ensuremath{ q\bar{q}}} \newc{\ppbar}{\ensuremath{ p\bar{p}}}
\newc{\bbbar}{\ensuremath{ b\bar{b}}} \newc{\ttbar}{\ensuremath{ t\bar{t}}}
\newc{\ffbar}{\ensuremath{ f\bar{f}}} \newc{\tautaubar}{\ensuremath{ \tau\bar{\tau}}}

\newc{\mchi}{\ensuremath{m_\neutone}}
\newc{\squark}{\ensuremath{\tilde{q}}}
\newc{\slepton}{\ensuremath{\tilde{l}}}
\newc{\gluino}{\ensuremath{\tilde{g}}} 
\newc{\mgluino}{\ensuremath{{m_{\gluino}}}}
\newc{\wino}{\ensuremath{\tilde{W}}} 
\newc{\mwino}{\ensuremath{{m_{\wino}}}}
\newc{\tone}{\ensuremath{{\tilde{t}_1}}}
\newc{\Hone}{\ensuremath{{\tilde{H}_{1}}}}
\newc{\Htwo}{\ensuremath{{\tilde{H}_{2}}}}
\newc{\Hhtwo}{\ensuremath{{H_{2}}}}
\newc{\qli}{\ensuremath{{\tilde{Q}_{i}}}}
\newc{\uri}{\ensuremath{{\tilde{u}_{i}}}}
\newc{\dri}{\ensuremath{{\tilde{d}_{i}}}}
\newc{\lli}{\ensuremath{{\tilde{L}_{i}}}}
\newc{\eri}{\ensuremath{{\tilde{e}_{i}}}}

\newc{\sthw}{\ensuremath{ \sin\theta_W}}              \newc{\cthw}{\ensuremath{\cos\theta_W}}
\newc{\tanthw}{\ensuremath{ \tan\theta_W}}              \newc{\cotthw}{\ensuremath{\cot\theta_W}}
\newc{\ssqthw}{\ensuremath{\sin^2 \theta_W}}
\newc{\msbar}{\ensuremath{\overline{MS}}} \newc{\drbar}{\ensuremath{\overline{DR}}}
\newc{\mtmtsmmsbar}{\ensuremath{ m_t(m_t)^{\msbar}_{{\mathrm{SM}}}}}
\newc{\mtmtsmdrbar}{\ensuremath{ m_t(m_t)^{\drbar}_{{\mathrm{SM}}}}}
\newc{\mtmtmssmdrbar}{\ensuremath{ m_t(m_t)^{\drbar}_{{\mathrm{SUSY}}}}}
\newc{\mbmbmsbar}{\ensuremath{ m_b(m_b)^{\msbar} }}
\newc{\mbmbsmmsbar}{\ensuremath{ m_b(m_b)^{\msbar}_{{\mathrm{SM}}}}}
\newc{\mbmzsmmsbar}{\ensuremath{ m_b(\mz)^{\msbar}_{{\mathrm{SM}}}}}
\newc{\mbmzsmdrbar}{\ensuremath{ m_b(\mz)^{\drbar}_{{\mathrm{SM}}}}}
\newc{\mbmzmssmdrbar}{\ensuremath{ m_b(\mz)^{\drbar}_{{\mathrm{SUSY}}}}}
\newc{\mtaumzsmmsbar}{\ensuremath{ m_{\tau}(\mz)^{\msbar}_{{\mathrm{SM}}}}}
\newc{\mtaumzsmdrbar}{\ensuremath{ m_{\tau}(\mz)^{\drbar}_{{\mathrm{SM}}}}}
\newc{\mtaumzmssmdrbar}{\ensuremath{ m_{\tau}(\mz)^{\drbar}_{{\mathrm{SUSY}}}}}
\newc{\alphasmzms}{\ensuremath{\alpha_s(M_Z)^{\overline{MS}}}}
\newc{\alphaimzms}[1]{\ensuremath{\alpha_{#1}(M_Z)^{\overline{MS}}}}
\newc{\alphaemmz}{\ensuremath{\alpha_{\mathrm{em}}(M_Z)^{\overline{MS}}}}

\newc{\mzero}{\ensuremath{{m_0}}}
\newc{\mhalf}{\ensuremath{ m_{1/2}}}
\newc{\tanb}{\ensuremath{\tan\beta}}
\newc{\azero}{\ensuremath{ A_0}}
\newc{\signmu}{\ensuremath{\rm{sgn}\,\mu}}
\newc{\atau}{\ensuremath{{A_{\tau}}}}
\newc{\mueff}{\ensuremath{\mu_{\rm{eff}}}}
\newc{\lam}{\ensuremath{{\lambda}}}
\newc{\kap}{\ensuremath{{\kappa}}}
\newc{\alam}{\ensuremath{{A_{\lambda}}}}
\newc{\akap}{\ensuremath{{A_{\kappa}}}}
\newc{\hs}{\ensuremath{ H_s}}      
\newc{\mhs}{\ensuremath{ m_{H_s}}} 
\newc{\mgut}{\ensuremath{ M_{\rm GUT}}}
\newc{\mvl}{\ensuremath{ M_{\rm VL}}}
\newc{\gut}{\ensuremath{{\rm GUT}}}
\newc{\mplanck}{\ensuremath{ M_{\rm P}}}      \newc{\mpl}{\ensuremath{ M_{\rm Pl}}}
\newc{\msusy}{\ensuremath{ M_{\rm SUSY}}}      \newc{\ms}{\ensuremath{ M_{\rm S}}}
 \newc{\hu}{\ensuremath{ H_u}}       \newc{\hd}{\ensuremath{ H_d}}
 \newc{\mhu}{\ensuremath{ m_{H_u}}}       \newc{\mhd}{\ensuremath{ m_{H_d}}}
 \newc{\mhuew}{\ensuremath{ m^{\ast}_{H_u}}}       \newc{\mhdew}{\ensuremath{ m^{\ast}_{H_d}}}
 \newc{\mhuewsq}{\ensuremath{ m^{\ast\, 2}_{H_u}}}       \newc{\mhdewsq}{\ensuremath{ m^{\ast\, 2}_{H_d}}}
 \newc{\mhl}{\ensuremath{m_\hl}} 
 \newc{\mhone}{\ensuremath{m_{h_1}}} 
 \newc{\mhtwo}{\ensuremath{m_{h_2}}} 
 \newc{\mhi}{\ensuremath{m_{\tilde{h}}}} 
 \newc{\mul}{\ensuremath{m_{\tilde{u}_L}}} 
 \newc{\mtone}{\ensuremath{m_{\tilde{t}_1}}} 
 \newc{\ma}{\ensuremath{m_A}} 
 \newc{\mH}{\ensuremath{m_H}} 
 \newc{\maone}{\ensuremath{m_{a_1}}} 
 \newc{\matwo}{\ensuremath{m_{a_2}}}
 \newc{\hone}{\ensuremath{h_1}}
 \newc{\htwo}{\ensuremath{h_2}}
 \newc{\aone}{\ensuremath{a_1}}
 \newc{\atwo}{\ensuremath{a_2}}
 \newc{\mqthree}{\ensuremath{m_{\tilde{Q}_3}^2}}
 \newc{\muthree}{\ensuremath{m_{\tilde{u}_3}^2}}
 \newc{\mqli}{\ensuremath{m_{\tilde{Q}_{i}}}}
 \newc{\muri}{\ensuremath{m_{\tilde{u}_{i}}}}
 \newc{\mdri}{\ensuremath{m_{\tilde{d}_{i}}}}
 \newc{\mlli}{\ensuremath{m_{\tilde{L}_{i}}}}
 \newc{\meri}{\ensuremath{m_{\tilde{e}_{i}}}}
 \newc{\ts}{\ensuremath{T_{SUSY}}}

\newc{\sigsip}{\ensuremath{\sigma^{\rm SI}_{p}}}	\newc{\sigsin}{\ensuremath{\sigma^{\rm SI}_{n}}}
\newc{\sigsdp}{\ensuremath{\sigma^{\rm SD}_{p}}}	\newc{\sigsdn}{\ensuremath{\sigma^{\rm SD}_{n}}}
\newc{\sigsi}{\ensuremath{\sigma^{\rm SI}}}	\newc{\sigsd}{\ensuremath{\sigma^{\rm SD}}}
\newc{\abund}{\ensuremath{ \Omega h^2}}
\newc{\omegadm}{\ensuremath{ \Omega_{{\rm DM}}}}     \newc{\abunddm}{\ensuremath{ \Omega_{{\rm DM}} h^2}} 
\newc{\omegam}{\ensuremath{ \Omega_{{\rm m}}}}       \newc{\abundm}{\ensuremath{ \Omega_{{\rm m}} h^2}}
\newc{\omegab}{\ensuremath{ \Omega_{{\rm b}}}}	\newc{\abundb}{\ensuremath{ \Omega_{{\rm b}} h^2}}
\newc{\omegatot}{\ensuremath{ \Omega_{{\rm TOT}}}}
\newc{\omegacdm}{\ensuremath{ \Omega_{{\rm CDM}}}}   \newc{\abundcdm}{\ensuremath{ \Omega_{{\rm CDM}} h^2}}
\newc{\omegalambda}{\ensuremath{ \Omega_{\Lambda}}} \newc{\abundlambda}{\ensuremath{ \Omega_{\Lambda} h^2}}
\newc{\omegarad}{\ensuremath{ \Omega_{{\rm rad}}}}  \newc{\abundrad}{\ensuremath{ \Omega_{{\rm rad}} h^2}}
\newc{\rhocrit}{\ensuremath{ \rho_{\rm crit}}}
\newc{\rhochi}{\ensuremath{ \rho_{\chi}}}
\newc{\abunchi}{\ensuremath{\Omega_\chi h^2}}
\newc{\abundlsp}{\ensuremath{\Omega_{\rm LSP}h^2}}

\newc{\amu}{\ensuremath{ a_{\mu}}}        \newc{\amususy}{\ensuremath{ a_{\mu}^{\mathrm{SUSY}}}}
\newc{\amuexpt}{\ensuremath{ a_{\mu}^{\mathrm{expt}}}}        \newc{\amusm}{\ensuremath{ a_{\mu}^{\mathrm{SM}}}}
\newc\deltaamu{\ensuremath{\Delta a_{\mu}}} \newc{\deltaamususy}{\ensuremath{\delta a_{\mu}^{\mathrm{SUSY}}}}
\newc\gmtwo{\ensuremath{ (g-2)_{\mu}}} 
\newc{\deltagmtwomususy}{\ensuremath{\delta\left(g-2\right)_{\mu}^{\mathrm{SUSY}}}}
\newc{\deltagmtwomu}{\ensuremath{\delta\left(g-2\right)_{\mu}}}
\newc{\deltagmtwoe}{\ensuremath{\delta\left(g-2\right)_{e}}}
\newc\BR{\ensuremath{\rm BR}}
\newc\bsgamma{\ensuremath{ b\rightarrow s \gamma }}
\newc\bxsgamma{\ensuremath{\overline{B}\rightarrow X_{s}\gamma}}
\newc\brbsgamma{\ensuremath{\BR\left(\bsgamma\right)}}
\newc\brbxsgamma{\ensuremath{\BR\left(\bxsgamma\right)}}
\newc\bsmumu{\ensuremath{B_s\to\mu^+\mu^-}}
\newc\brbsmumu{\ensuremath{\BR\left(B_s\to\mu^+\mu^-\right)}}
\newc\bdmmumu{\ensuremath{\overline{B}_d\to\mu^+\mu^-}}
\newc\bbbarmix{\ensuremath{\overline{B}_s\mbox{-}B_s}}      
\newc\delmbs{\ensuremath{\Delta M_{B_s}}}
\newc{\butaunu}{\ensuremath{B_u \rightarrow \tau \nu}}
\newc{\brbutaunu}{\ensuremath{\BR\left(B_u \rightarrow \tau \nu\right)}}

\newcommand*{\reftable}[1]{Table~\ref{#1}}         
       
\newcommand*{\reffig}[1]{Fig.~\ref{#1}}
 
        \newcommand*{\refeq}[1]{Eq.~(\ref{#1})}
     \newcommand*{\refsec}[1]{Sec.~\ref{#1}}

\newcommand*{\refeqs}[2]{Eqs.~(\ref{#1})-(\ref{#2})}
\newcommand*{\neutone}{\ensuremath{\tilde{\chi}^0_1}}

\newcommand*{\madgr}{\texttt{MadGraph5\_aMC@NLO}}

\newcommand*{\chmt}{\texttt{CheckMATE}}

\let\oldcite\cite
\renewcommand*{\cite}{~\oldcite}

\newcommand*{\hl}{\ensuremath{h}}

\newcommand*{\smgaux}{SU(3)$_{\textrm{c}}\times$SU(2)$_L\times$U(1)$_Y\times$U(1)$_X$}
\newcommand*{\eps}{\ensuremath{\epsilon}}